\documentclass[preprint,aps,pra,showpacs,floatfix]{revtex4-1}
\usepackage{bm}
\usepackage{dcolumn}
\usepackage{graphicx}
\begin{document}
\title{Electron elastic scattering and low-frequency bremsstrahlung on $A$@C$_{60}$: A model static approximation}
\author{V. K. Dolmatov}
\author{C. Bayens}
\author{M. B. Cooper}
\author{M. E. Hunter}
\affiliation{Department of Physics
and Earth Science, University of North Alabama, Florence, Alabama 35632, USA}

\begin{abstract}
Electron elastic-scattering phase shifts and cross sections along with the differential and total cross sections and polarization of low-frequency bremsstrahlung upon low-energy electron collision with endohedral fullerenes  $A$@C$_{60}$ are theoretically scrutinized versus the nature, size and spin of the encapsulated atom $A$. The case-study-atoms $A$ are  N, Ar, Cr, Mn, Mo, Tc, Xe, Ba, and Eu. They are thoughtfully picked out of different rows of the periodic table. The study is performed in the framework of a model static approximation. There, both the encapsulated atom $A$ and C$_{60}$ cage are regarded as non-polarizable targets. The  C$_{60}$ cage is modeled by an attractive spherical annular-potential well.
The study provides the most complete \textit{initial} understanding of how the processes of interest might evolve upon electron collision with various $A$@C$_{60}$. Calculated results identify the most
 interesting and/or useful future measurements or more rigorous calculations to perform.
\end{abstract}
\maketitle

\section{Introduction}
Electron elastic scattering and bremsstrahlung (a process of emission of radiation upon collision of electrons with matter) on quantum targets are important fundamental phenomena of nature with significance to both the basic and applied sciences and technologies. Yet, to date, the knowledge on these phenomena upon electron collision with such important quantum targets as endohedral fullerenes $A$@C$_{60}$  is largely lacking. Endohedral fullerenes,
(also referred to, interchangeably, as \textit{endohedral atoms} or just \textit{fullerenes} in the present paper) are nano-structure formations where an atom $A$ is encapsulated inside the hollow interior of a C$_{60}$ fullerene. They are relatively novel and important objects of intense modern studies.
In fact, the authors are aware of only one published work on the subject of low-energy electron elastic scattering off $A$@C$_{60}$ \citep{e+A@C60}. Also, to the authors' best knowledge, there seems to be an absence of
a study of low-frequency bremsstrahlung by low-energy electrons scattered off $A$@C$_{60}$ endohedral fullerenes. It is the ultimate aim of the present paper (a) to get a broader insight into properties of low-energy electron elastic scattering off $A$@C$_{60}$, (b) to provide the initial insight into features of electron low-frequency bremsstrahlung on $A$@C$_{60}$, and (c) to explore to a greater extent how said properties and features might evolve with changing the size, softness, and spin of the encapsulated atom. To meet this goal, the authors pick typical representatives of atoms from the family of noble gases (namely, N, Ar, and Xe), $3d$ and $4d$ transition-metals (Cr, Mn, Mo, and Tc), alkaline (Ba), and rare-earth (Eu) elements of the periodic table. As a result, the basic features as well as characteristic similarities and discrepancies of electron elastic scattering and low-frequency bremsstrahlung on various endohedral fullerenes $A$@C$_{60}$ are revealed, interpreted, and detailed. The quantities subjected to this study are the low-energy electron elastic-scattering phase shifts and cross sections as well as the total cross sections, dipole angular-asymmetry distributions, and the angle-resolved degree of polarization of low-frequency bremsstrahlung.

The present study also has a significance which is independent of its direct applicability to endohedral fullerenes. This is because it falls into a mainstream of intense modern studies where numerous aspects of the structure and spectra of atoms under various kinds of confinements are being attacked from many different angles by
research teams world-wide. This has resulted in a huge array of unraveled effects and data being accumulated in a large number of publications to date, see reviews
\citep{Jaskolski,Buchachenko,RPC,AQC57,AQC58,Kalidas14} (and references therein). There, one finds a wealth of information
on properties of single-electron, two-electron
and many-electron atoms confined  by impenetrable spherical, spheroidal, as well as open boundary potentials
 (e.g., see review papers in \citep{AQC57} by Aquino, p.~$123$; Laughlin, p.~$203$; Cruz, p.~$255$;   Garza and Vargas, p.~$241$),
oscillator potentials (e.g.,  Patil and Varshni \citep{AQC57}, p.~1), potentials limited by conoidal boundaries (Ley-Koo \citep{AQC57}, p~.79), Debye potentials
(Sil, Canuto, and Mukherjee \citep{AQC58}, p.~115),
fullerene-cage potentials (Dolmatov \citep{AQC58}, p.13; Charkin et.\,al.~\citep{AQC58}, p.69, Amusia et.\,al. \citep{C60-}), potential with dihedral angles (Ley-Koo and Sun \citep{Kalidas14}, p.~1), \textit{etc}. The present study adds new basic knowledge
to the existing collection of atomic properties under confinement.

The interaction of radiation and charged particles with endohedral atoms is a complicated multifaceted process. This is in view of a great variety of various effects that contribute to the process. It is, therefore, both desirable and important to understand how each of the ``facets'' contributes to, and results in this or that effect in, the processes of interest, rather than to get only the cumulative result. In the present paper, we expose to light the impact of a ``static facet'' on $e + A@{\rm C}_{60}$ elastic scattering and bremsstrahlung. This is achieved by considering these processes in the framework of an approximation referred to as the
\textit{model static approximation}
in the present paper. In this approximation, the C$_{60}$ cage is modeled by an attractive spherical annular-potential well $U_{\rm c}(r)$ of certain inner radius $r_{0}$, width $\Delta$, and depth $U_{0}$. The C$_{60}$ cage, thus, is regarded as a non-polarizable target. The encapsulated atom $A$ is positioned at the center of the potential $U_{\rm c}(r)$  and is regarded as a non-polarizable target as well. The potential of $A$@C$_{60}$ is defined as the sum of the potential $U_{\rm c}(r)$ and non-local Hartree-Fock (HF) potential of the encapsulated atom $A$. The corresponding HF equation is then solved in order to determine the wavefunctions and electron elastic-scattering phase shifts upon $e + A@{\rm C}_{60}$ collision. Note that this approximation, where the C$_{60}$ is modeled by the the potential $U_{\rm c}(r)$ with the atom $A$ being at the center of the potential, has been used for the study of the interaction of photons and charged particles with endohedral fullerenes $A$@C$_{60}$ on numerous occasions to now, see, e.g.,  \citep{e+A@C60,Pushka,RPC,e+Xe,Ludlow,@TD14} (and references therein). Also, the replacement of the C$_{60}$ cage by the same potential
$U_{\rm c}(r)$ was employed in work \citep{Winstead} for the study of electron elastic scattering off empty C$_{60}$ as well. In the same work, the study of $e + {\rm C}_{60}$ scattering was paralleled by the
 calculation performed in the framework of a sophisticated \textit{ab initio} molecular-Hartree-Fock approximation combined with the Schwinger multichannel scattering theory.
The work \citep{Winstead} provided a thorough, detailed comparison of calculated results for the $e + {\rm C}_{60}$ scattering phase shifts as well as partial and total elastic-scattering cross sections obtained in the frameworks of these two approximations. A reasonable qualitative, and even semi-quantitative, agreement between some of the most prominent features of $e + {\rm C}_{60}$ elastic scattering, predicted by the two calculations, was demonstrated. Such agreement speaks in favor of the overall usability of the $U_{\rm c}$-model-potential approximation to electron-fullerene collision.

In the present work, the electron collision energy $\epsilon$ is assumed to be sufficiently small
($\epsilon \le 15$ eV). At such energies, the electron wavelength $\lambda > 3$ $\AA$. It, thus, exceeds noticeably the bond length $D \approx 1.44$ $\AA$ between the carbon atoms in C$_{60}$. Correspondingly, the incoming electrons will ``see'' the C$_{60}$ cage as a homogeneous rather than ``granular'' cage. This justifies the modeling of the C$_{60}$ cage by a smooth potential, in general, such as the above introduced potential $U_{\rm c}(r)$, in particular. Furthermore, in the present work, the emphasis is on low-frequency bremsstrahlung, $\omega \rightarrow 0$. In the latter case, (a) the bremsstrahlung phenomenon can easily be attacked in the framework of a low-frequency approximation \citep{Johnson} and (b) the contribution of a tricky ``polarization bremsstrahlung'' amplitude \citep{ATOM,Astapenko,Solov'yov} (and references therein) can be safely excluded from the study. (The ``polarization bremsstrahlung'' amplitude is the amplitude of the photon emission by a target during its dynamical polarization by an incoming electron).

In summary, the model static approximation employed in the present paper for the study of both low-energy electron elastic scattering and low-frequency bremsstrahlung upon $e+A@{\rm C}_{60}$ collision is overall reasonable.
Its drawback is the omission of accounting for electron correlation in the $e+A@{\rm C}_{60}$ system (or, which is the same, the omission of polarization of the C$_{60}$ cage and/or atom $A$ by an incident electron). However, first, a thorough description of electron scattering and bremsstrahlung on a multielectron target is too challenging for theorists even with regard to a free atom, not to mention a $A$@C$_{60}$ target; the development of a a corresponding comprehensive theory is for future years. Second, in order to understand, interpret, and appreciate the impacts of correlation and other omitted effects on $e + A@{\rm C}_{60}$ elastic scattering and bremsstrahlung one \textit{does} need to know how the processes develop without accounting for such effects. The present study provides researchers exactly with such knowledge. Moreover, the model static approximation allows one to uncover characteristic properties of the investigated phenomena which \textit{do} exist without regard for
details of bonding between the $60$ carbon atoms of the C$_{60}$ cage, \textit{etc}. In a sense, the present work unveils some of the most basic intrinsic
properties of low-energy electron elastic scattering and electron low-frequency bremsstrahlung off $A@{\rm C}_{60}$ fullerenes. It identifies the most
 interesting and/or useful future measurements or more rigorous calculations to be performed in order to advance this field of study.

\section{Theory}

In the present work, the C$_{60}$ cage is modeled by a spherical annular-potential well, $U_{\rm c}(r)$:
\begin{eqnarray}
U_{\rm c}(r)=\left\{\matrix {
-U_{0}, & \mbox{if $r_{0} \le r \le r_{0}+\Delta$} \nonumber \\
0 & \mbox{otherwise.} } \right.
\label{SWP}
\end{eqnarray}
Here, $r_{0}$, $\Delta$, and $U_{0}$ are the inner radius, thickness, and depth of the potential well, respectively;
their magnitudes are borrowed from Ref.~\citep{Winstead}. Namely,
$\Delta = 2.9102$ $a_{0}$ ($a_{0}$ being the first Bohr radius of the hydrogen atom), $r_{0} = R_{\rm c} - (1/2)\Delta =5.262$ $a_{0}$ ($R_{\rm c}=6.7173$ $a_{0}$ being the radius of the C$_{60}$ skeleton),
and $U_{0} = 7.0725$ eV (found by matching the electron affinity $EA=-2.65$ eV of C$_{60}$ with the assumption that the orbital momentum of the $2.65$-eV-state is $\ell =1$). These values
of the adjustable parameters are most consistent with the corresponding observations.

Next, the wavefunctions $\psi_{n \ell m_{\ell} m_{s}}({\bm r}, \sigma)=r^{-1}P_{nl}(r)Y_{l m_{\ell}}(\theta, \phi) \chi_{m_{s}}(\sigma)$
and binding energies $\epsilon_{n l}$ of atomic electrons
($n$, $\ell$,  $m_{\ell}$ and $m_{s}$ is the standard set of quantum numbers of an electron in a central field, $\sigma$ is the electron spin coordinate) are the solutions of a system of the ``endohedral''
HF equations [in atomic units (a.~u.~)]:
\begin{eqnarray}
&&\left[ -\frac{\Delta}{2} - \frac{Z}{r} +U_{\rm c}(r) \right]\psi_{i}
({\bm x}) + \sum_{j=1}^{Z} \int{\frac{\psi^{*}_{j}({\bm x'})}{|{\bm
x}-{\bm x'}|}} \nonumber \\
 && \times[\psi_{j}({\bm x'})\psi_{i}({\bm x})
- \psi_{i}({\bm x'})\psi_{j}({\bm x})]d {\bm x'} =
\epsilon_{i}\psi_{i}({\bm x}).
\label{eqHF}
\end{eqnarray}
Here, $Z$ is the nuclear charge of the atom, ${\bm x} \equiv ({\bm r}, \sigma)$, and the integration over ${\bm x}$ implies both the integration over ${\bm r}$ and summation over
$\sigma$. Eq.~(\ref{eqHF}) differs from the ordinary HF equation for a free atom by the presence of the $U_{\rm c}(r)$ potential in the equation. This equation is first solved in order to calculate the electronic ground-state wavefunctions of the encapsulated atom. Once the electronic ground-state wavefunctions are determined, they are plugged back into
 Eq.~(\ref{eqHF}) in place of the $\psi_{j}({\bm x'})$ and $\psi_{j}({\bm x})$ functions in order to calculate the electronic wavefunctions of scattering-states $\psi_{i}({\bm x})$ and their radial parts $P_{\epsilon_{i}\ell_{i}}(r)$. Corresponding electron elastic-scattering phase shifts $\delta_{\ell}(k)$
  are then determined by
 referring to $P_{k\ell}(r)$ at large $r$ \citep{Landau}:
\begin{eqnarray}
P_{k\ell}(r) \rightarrow \sqrt{\frac{2}{\pi}}\sin\left(k r -\frac{\pi\ell}{2}+\delta_{\ell}(k)\right).
\label{P(r)}
\end{eqnarray}
Here, $k$ is the electron's wavenumber [$k \equiv |{\bm k}|= (2m\epsilon / \hbar^{2})^{1/2}$, ${\bm k}$ and $m$ being the electron's wavevector and mass, respectively],
$P_{k\ell}(r)$ is normalized to $\delta(k-k')$, where $k$ and $k'$ are the wavenumbers of the incident and scattered electrons, respectively.
The total electron elastic-scattering cross section $\sigma_{\rm el}(\epsilon)$ is then found in accordance with the well-known
formula for electron scattering by a central-potential field \citep{Landau}:
 \begin{eqnarray}
 \sigma_{\rm el}(k)= \frac{4\pi}{k^2}\sum^{\infty}_{\ell=0}(2\ell+1)\sin^{2}\delta_{\ell}(k).
 \label{sigma}
 \end{eqnarray}

A differential cross section $d\sigma(\omega)$ of bremsstrahlung into the frequency interval $d\omega$, the direction of the photon momentum ${\bf p}_{ph}=\hbar{\bm q}$ into the solid angle $d\Omega_{\bm q}$, and
the direction of the momentum ${\bf p'}=\hbar{\bm k'}$  of a scattered electron into $d\Omega_{\bm k'}$ is defined as follows \citep{Sobelman}:
\begin{eqnarray}
d\sigma(\omega) = &&\frac{
{m^{2}}{e^{2}}q^{3} k'
}
{
(2\pi)^{4}{\rm \hbar^{3}}k
} \nonumber \\
&& \times \left|{\bm {\hat e}_{q}}\int(\psi_{\bm k}^{+})^{*}{\bm r}\psi_{\bm k'}^{-}d{\bm r}\right|^{2}d\omega d\Omega_{\bm q} d\Omega_{\bm k'}.
\label{BS1}
\end{eqnarray}
Here, $\hbar q {\rm c} =\hbar\omega= \frac{\hbar^2 k^2}{2 m}-\frac{\hbar^2 k'^2}{2 m}$,  where $\rm c$ is the speed of light, $e$ is the electronic charge,
$\bm k'$ is the wavevector of the scattered electron, $\hat{\bm e}_{q}$ is the unit vector of the photon polarization, and $\psi_{\bm k}^{\pm}$ are the wavefunctions of
the incident and scattered electrons, respectively:
\begin{eqnarray}
\psi_{\bm k}^{\pm}({\bm r}) = && \frac{(2\pi)^{3/2}}{k} \sum_{\ell, \mu}{\rm i}^{\ell}\exp{[\pm{\rm i}\delta_{\ell}(k)]} \nonumber \\
&&  \times Y_{\ell m_{\ell}}^{*}(\theta_{\bm k},\phi_{\bm k})Y_{\ell m_{\ell}}(\theta_{\bm r},\phi_{\bm r})\frac{P_{k\ell}(r)}{r}.
\label{EqPSI}
\end{eqnarray}
In the above equation, $\theta_{\bm k}$ and $\phi_{\bm k}$ are the spherical angles of the electron wavevector $\bm k$, whereas $\theta_{\bm r}$ and $\phi_{\bm r}$ are the spherical angles of the
electron position vector $\bm r$.

 Let us position the origin of a rectangular $XYZ$-system of coordinates on the encapsulated atom $A$. Let us assume that the momentum ${\bm p} =\hbar{\bm k}$ of an incident electron lies along the $Z$-axis, pointing in its positive direction. Furthermore, in the final state of the system, let us measure the directions of both the momentum ${\bm p}_{\rm ph}=\hbar {\bm q}$ of an emitted photon and its polarization vector $\hat{\bm e}_{q}$. The
vector  $\hat{\bm e}_{q}$ will be determined relative to a $({\bm p},{\bm p}_{ph})$-plane, being either parallel (${\bm e_{q}}_{\|}$) or perpendicular (${\bm e_{q}}_{\bot}$) to the plane. Then, with the help of Eq.~(\ref{BS1}), one can determine
the corresponding differential cross sections $d\sigma^{\bot}/d\omega d\Omega_{\bm q}$ and $d\sigma^{\|}/d\omega d\Omega_{\bm q}$  into the unit intervals of $\omega$ and $\Omega_{\bm q}$:
\begin{eqnarray}
\frac{d\sigma^{\bot}}{d\omega d\Omega_{\bm q}} = \frac{1}{8\pi}\frac{d\sigma}{d\omega}\left[1-\frac{1}{2}\beta(\omega)\right],
\label{EqPerp}
\end{eqnarray}
\begin{eqnarray}
\frac{d\sigma^{\|}}{d\omega d\Omega_{\bm q}} = \frac{1}{8\pi}\frac{d\sigma}{d\omega}\left\{1 + \frac{1}{2}\beta(\omega)[1-2P_{2}(\cos\theta)]\right\}.
\label{Eq||}
\end{eqnarray}
Here, $P_{2}(\cos\theta)$ is the Legendre polynomial of the second order, $\theta$ is the angle between the $Z$ axis and the photon momentum ${\bm p}_{ph}$, $d\sigma/d\omega$ is the bremsstrahlung angle-integrated cross section (or, interchangeably, the spectral density of bremsstrahlung) \citep{Sobelman}, and $\beta(\omega)$ is the angular-asymmetry parameter of bremsstrahlung:
\begin{eqnarray}
\frac{d\sigma}{d\omega} = &&\frac{8\pi^2}{3} \frac{m^2 \hbar^{4} \alpha^{3}}{\rm e^4}\frac{\omega^{3}}{p\prime p^{3}} \nonumber \\
&& \times\sum_{\ell=0}^{\infty}\left[\ell D_{\ell-1}^{2}(p) + (\ell+1)D_{\ell+1}^{2}(p)\right],
\label{EqB1}
\end{eqnarray}
\begin{eqnarray}
\beta(\omega) = &&\left[\sum_{\ell=0}^{\infty}\left(\ell D_{\ell-1}^{2} + (\ell+1)D_{\ell+1}^{2}\right)\right]^{-1} \nonumber \\
&&\times
\sum_{\ell=0}^{\infty} (2\ell+1)^{-1}\left[(\ell+1)(\ell+2)D_{\ell+1}^{2}\right. \nonumber \\
&&\left.+\ell(\ell-1)D_{\ell-1}^{2} -6\ell(\ell+1) D_{\ell+1} D_{\ell-1} \right. \nonumber \\
&&\left.\times\cos\left(\delta_{\ell+1} - \delta_{\ell-1}\right )\right ].
\label{Eqbeta}
\end{eqnarray}
Here, $\alpha$ is the fine structure constant and $D_{\ell\pm 1}$ is the bremsstrahlung dipole  amplitude:
\begin{eqnarray}
D_{\ell\pm 1}= \int_{0}^{\infty}P_{k',\ell \pm 1} r P_{k,\ell}(r)dr.
\label{EqD}
\end{eqnarray}

To determine the differential cross section $d\sigma/d\omega d\Omega_{\bm q}$ of unpolarized bremsstrahlung, one adds Eqs.~(\ref{EqPerp}) and (\ref{Eq||}) together and arrives at the known formula (see, e.g., \citep{Balt}):
\begin{eqnarray}
\frac{d\sigma}{d\omega d\Omega_{\bm q}} = \frac{1}{4\pi}\frac{d\sigma}{d\omega}\left [1 - \frac{1}{2}\beta(\omega)P_{2}(\cos\theta)\right ],
\end{eqnarray}
where the parameter $\beta(\omega)$ is given by the same Eq.~(\ref{Eqbeta}).

Next, the parameter of the degree of the bremsstrahlung's polarization, $\zeta_{3}$ (known as the Stokes third parameter), defined as the ratio of the difference between
$d\sigma^{\bot}(\omega)/d\omega d\Omega$ and $d\sigma^{\|}(\omega)/d\omega d\Omega$ to their sum, takes the following form:
\begin{eqnarray}
\zeta_{3}(\theta)=\frac{\beta[1-P_{2}(\cos\theta)]}{2-\beta P_{2}(\cos\theta)}.
\label{Eqzeta}
\end{eqnarray}

In the framework of the low-frequency bremsstrahlung approximation ($\omega \rightarrow 0$), utilized in the present paper, $\epsilon_{\rm i}\approx \epsilon_{\rm f}$ ($\epsilon_{\rm i}$
and $\epsilon_{\rm f}$ are the initial and final electron energy, respectively). In this case, the functions
$P_{k,\ell}(r)$ and $P_{k'\ell\pm 1}$ in Eq.~(\ref{EqD}) can \citep{Johnson} be replaced by their asymptotic
forms, Eq.~(\ref{P(r)}). Correspondingly, one readily obtains, see, e.g., \citep{Balt,Johnson}:
\begin{eqnarray}
D_{\ell\pm 1}(\omega)|_{\omega \rightarrow 0} = \pm \frac{1}{\pi}\left (\frac{p}{m\omega}\right)^{2}\sin[\delta_{\ell}(p) - \delta_{\ell\pm 1}(p)].
\label{EqD'}
\end{eqnarray}

As was noted in the previous section, some of the encapsulated atoms of interest  are N, Cr, Mn, Mo, Tc, and Eu.  Theses are high-spin atoms, owing to one or two semifilled subshells in their ground-state configurations:
N(...$2p^{3}$, $^{4}S$) (with the single semifilled subshell $2p^{3}$),
Cr(...$\rm 3d^{5}$$\rm 4s^{1}$, $^{7}$S) (with the two semifilled subshells $3d^{5}$ and $4s^{1}$),
Mn(...$\rm 3d^{5}$$\rm 4s^{2}$, $^{6}$S) (with the single semifilled subshell $3d^{5}$),
Mo(...$\rm 4d^{5}$$\rm 5s^{1}$, $^{7}$S) (with the two semifilled subshells $4d^{5}$ and $5s^{1}$),
Tc(...$\rm 4d^{5}$$\rm 5s^{2}$, $^{6}$S) (with the single semifilled subshell $4d^{5}$),
and Eu(...$\rm 4f^{7}$$\rm 6s^{2}$, $^{8}$S) (with the single semifilled subshell $4f^{7}$).
Atoms with open as well as semifilled subshells require a special approach to the calculation of their structure and spectra. A convenient, effective
 theory to calculate the structure of a semifilled shell atom  is the ``spin-polarized'' Hartree-Fock (SPHF) approximation developed by Slater \citep{Slater}.  The quintessence of SPHF is as follows. It accounts for the fact that spins of all
electrons in the semifilled subshell(s) of the atom (e.g., in the $\rm 3d^{5}$$\uparrow$ and $\rm 4s^{1}$$\uparrow$  subshells in the Cr atom ) are co-directed, in accordance with Hund's rule, say, all pointing upward. This results in splitting of each of other
closed ${n\ell}^{2(2\ell+1)}$ subshells in the atom into two semifilled subshells of opposite spin orientations, ${n\ell}^{2\ell+1}$$\uparrow$ and ${n\ell}^{2\ell+1}$$\downarrow$. This is in view of
 the presence of
exchange interaction between $nl$$\uparrow$ electrons with only spin-up electrons in the original spin-unpaired semifilled
subshell(s) of the atom (like the $\rm 3d^{5}$$\uparrow$ and $\rm 4s^{1}$$\uparrow$  subshells in the Cr atom) but  absence of such for $nl$$\downarrow$ electrons. Thus, the SPHF configurations of the picked out
semifilled-subshell atoms  are as follows:\\\\
N(...${2s}^{1}$$\uparrow$${2s}^{1}$$\downarrow$${2p}^{3}$$\uparrow$, $^{4}$S),\\
Cr(...${3p}^{3}$$\uparrow$${3p}^{3}$$\downarrow$${3d}^{5}$$\uparrow$${4s}^{1}$$\uparrow$, $^{7}$S),\\
Mo(...${4p}^{3}$$\uparrow$${4p}^{3}$$\downarrow$${4d}^{5}$$\uparrow$${5s}^{1}$$\uparrow$, $^{7}$S),\\
Mn(...${3p}^{3}$$\uparrow$${3p}^{3}$$\downarrow$${3d}^{5}$$\uparrow$${4s}^{1}$$\uparrow$$4s^{1}$$\downarrow$, $^{6}$S),\\
Tc(...${4p}^{3}$$\uparrow$${4p}^{3}$$\downarrow$${4d}^{5}$$\uparrow$${5s}^{1}$$\uparrow$$5s^{1}$$\downarrow$, $^{6}$S),\\
Eu(...${4d}^{5}$$\uparrow$${4d}^{5}$$\downarrow$${4f}^{7}$$\uparrow$${6s}^{1}$$\uparrow$$6s^{1}$$\downarrow$, $^{8}$S).\\

 SPHF equations for the ground-state, bound excited-states and scattering-states of a semifilled shell atom differ from ordinary HF equations for closed shell atoms by accounting for exchange interaction only between electrons with the same spin orientation
($\uparrow$, $\uparrow$ or $\downarrow$, $\downarrow$). To date, SPHF  has successfully been extended to studies of electron elastic scattering off isolated semifilled shell atoms in a number of works \citep{e+Mn,e+Cr,Remeta} (and references therein).
In the present paper, SPHF is utilized for calculation  both of the atomic and scattering states of $A$@C$_{60}$ endohedral fullerenes, where $A$ is a semifilled shell atom.

\section{Results and discussion}

\subsection{Valence orbitals of the encapsulated atoms $A$ in $A$@C$_{60}$}
\begin{figure}[ht]
\center{\includegraphics[width=8cm]{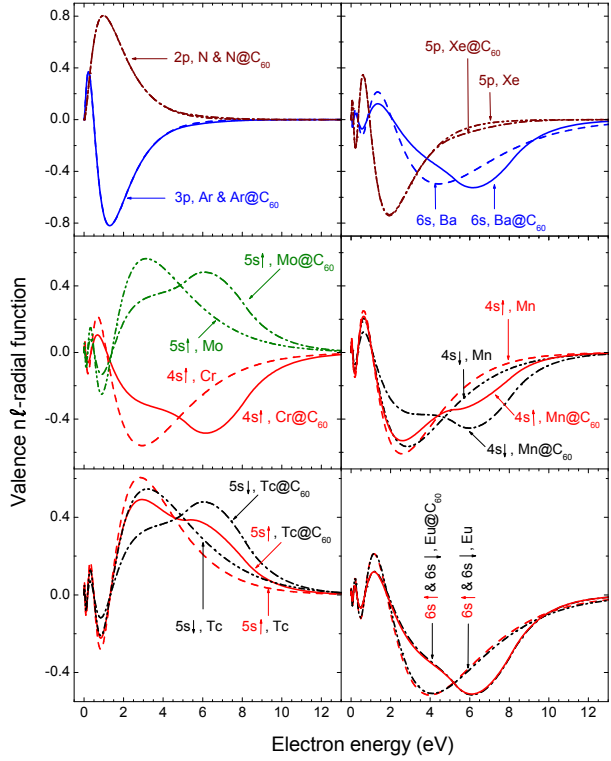}}
\caption{Calculated $P_{n\rm s\uparrow}(r)$ and $P_{n\rm s\downarrow}(r)$ radial functions (in atomic units) of the valence subshells of closed-shell Ar@C$_{60}$, Xe@C$_{60}$, and Ba@C$_{60}$ atoms \citep{e+A@C60}, as well as
semifilled-shell Cr@C$_{60}$,
Mn@C$_{60}$, Mo@C$_{60}$, Tc@C$_{60}$, and Eu@C$_{60}$ \citep{JPC&VKD15} atoms, and N@C$_{60}$ versus those of their free counter-parts, as marked; the spatial region $5.262 < r < 8.17$ a.u.
belongs to the wall of the C$_{60}$ cage.}
\label{FigWF}
\end{figure}

The impact of the C$_{60}$ cage on the valence orbitals of the encapsulated atoms of interest is illustrated by Fig.~\ref{FigWF}.

Note that the $2p^{3}$$\uparrow$ valence orbital of N@C$_{60}$, as well as the $3p^{6}$ valence orbital of Ar@C$_{60}$, practically coincide with respective orbitals of free N and Ar. Even the $5p^{6}$ valence orbital of a bigger Xe atom is only insignificantly altered upon its encapsulation inside of the C$_{60}$ cage. Therefore, these atoms are referred to as the ``compact'' atoms in the present paper. In contrast, the valence orbitals
of the Cr, Mn, Mo, Tc, Ba, and Eu atoms are significantly drawn into the potential well, i.e., into the region of the wall of C$_{60}$. These atoms are to be referred to as the ``soft'' atoms.

Next, note that the $4s$$\downarrow$-orbital of Mn is drawn into the C$_{60}$ wall noticeably stronger than the $4s$$\uparrow$-orbital. Similar difference emerges between the  $5s$$\uparrow$ and $5s$$\downarrow$ orbitals
of Tc as well.
This induces the transfer of a noticeable part of primarily the \textit{spin-down} electron density from the encapsulated atom to the C$_{60}$ cage. Correspondingly, the C$_{60}$ cage becomes, as it were, ``charged'' by a \textit{spin-down} electron density. This effect was originally spotted in Mn@C$_{60}$ \citep{e+A@C60}, where it was named the ``C$_{60}$-spin-charging effect''. Later, it was
detailed on a more extensive scale with an eye on the register of a quantum computer in \citep{JPC&VKD15}. In contrast to Mn@C$_{60}$ and Tc@C$_{60}$, the C$_{60}$ cage becomes \textit{spin-up} charged in Cr@C$_{60}$ and Mo@C$_{60}$. This is because of a significant \textit{spin-up} electron density drain from a $4s$$\uparrow$ spin-unpaired semifilled subshell  of Cr, or a $5s$$\uparrow$ spin-unpaired semifilled subshell of Mo, to the C$_{60}$ wall.
In contrast, the spin-dependent drain of the valence electron density does not take place in Eu@C$_{60}$. This is because the
$6s$$\uparrow$ and $6s$$\downarrow$ orbitals are drawn into the C$_{60}$ cage equally.  The latter, in turn, is because the $\rm 4f^{7}$$\uparrow$ semifilled  subshell of Eu lies much deeper relative to its $\rm 6s^{1}$$\uparrow$ and
$6s^{1}$$\downarrow$ subshells than the $nd^{5}$$\uparrow$
spin-unpaired semifilled  subshell of Mn and Tc relative their valence $ns$-subshells.  Correspondingly, the exchange interaction between the $\rm 4f$$\uparrow$ and $\rm 6s$$\uparrow$ electrons in Eu is negligible, and there is no exchange interaction between the $\rm 4f$$\uparrow$ and $\rm 6s$$\downarrow$ electrons. Hence,
there is practically no difference between the $6s$$\uparrow$ and $6s$$\downarrow$ orbitals of free or encapsulated Eu. As a result, the C$_{60}$ cage in Eu@C$_{60}$ is ``spin-neutral''.
Note that, as was argued in \citep{JPC&VKD15}, the C$_{60}$-spin-charging can affect the manipulation of spins in the corresponding $A$@C$_{60}$ systems and that it must inhibit, or at least render more complex, the operation of the register of a fullerene-based quantum computer \citep{Harneit}.

The above findings stir up one's mind by way of wonder: (a) how sensitive is electron elastic-scattering and bremsstrahlung to the \textit{size} of a \textit{compact} encapsulated atom?; (b) alternatively, how sensitive are these phenomena to the \textit{size} of a \textit{soft} encapsulated atom?; and (c) how sensitive are these phenomena to the \textit{spin} of an encapsulated atom?

The rest of the present work is motivated by the search for answers to the above questions.
\subsection{Electron collision with a  closed shell  $A$@C$_{60}$: $A= \rm Ar, Xe, Ba$}
\subsubsection{Electron elastic-scattering phase shifts}
\begin{figure}[ht]
\center{\includegraphics[width=8cm]{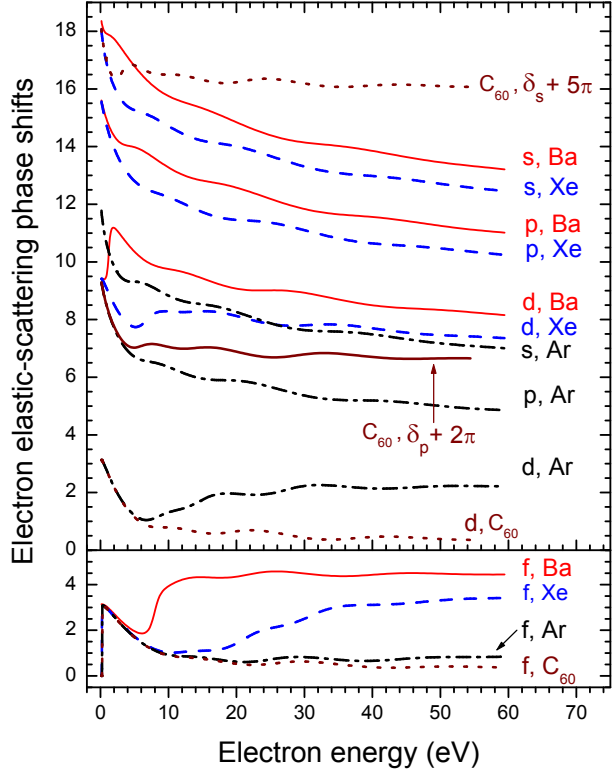}}
\caption{Calculated HF electron elastic-scattering phase shifts $\delta_{\ell}(\epsilon)$ (in units of radian) ($\ell = s, p, d$, and $f$) upon electron collision with Ar@C$_{60}$, Xe@C$_{60}$, Ba@C$_{60}$, and
 empty C$_{60}$, as marked.}
\label{FigArXeBa03}
\end{figure}

\begin{figure}[ht]
\center{\includegraphics[width=8cm]{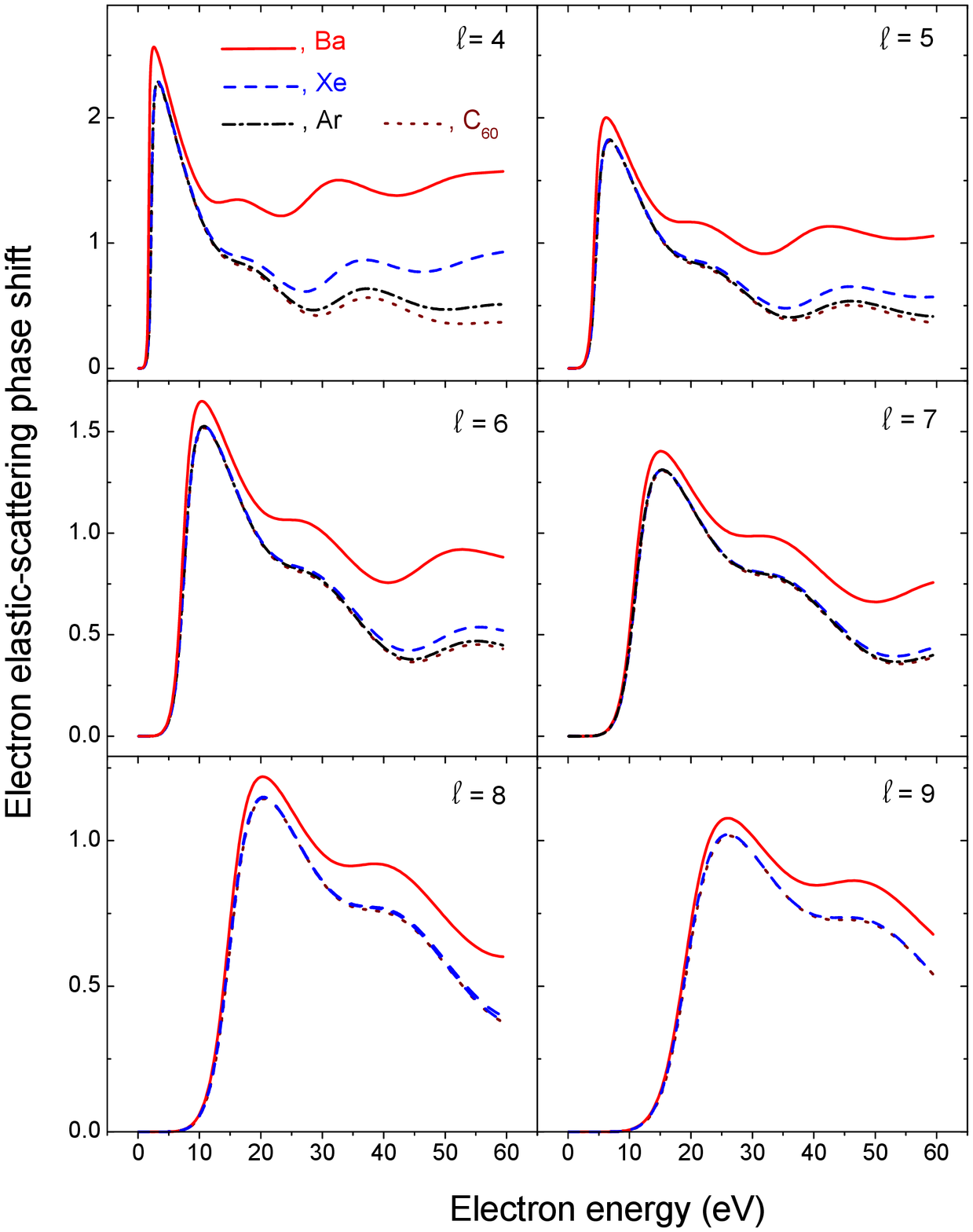}}
\caption{Calculated HF electron elastic-scattering phase shifts $\delta_{\ell}(\epsilon)$  (in units of radians) for $\ell \ge 4$ upon electron collision with Ar@C$_{60}$, Xe@C$_{60}$, Ba@C$_{60}$,
and empty C$_{60}$, as marked.}
\label{FigArXeBa49}
\end{figure}

Calculated electron elastic-scattering phase shifts $\delta_{\ell}(\epsilon)$ due to scattering off Ar@C$_{60}$, Xe@C$_{60}$, and Ba@C$_{60}$ and, for comparison, off empty C$_{60}$ are
depicted in Figs.~\ref{FigArXeBa03} and \ref{FigArXeBa49}.

Note that the depicted in Fig.~\ref{FigArXeBa03} phase shifts  $\delta_{\ell \le 3}^{A@{\rm C_{60}}}$ differ significantly from
the corresponding phase shifts $\delta_{\ell \le 3 }^{\rm C_{60}}$ for electron scattering off empty C$_{60}$, with some exceptions at low electron energies.

Additionally, interesting differences in $\delta_{\ell}^{A@{\rm C_{60}}}$s between different fullerenes emerge.
The differences are
mostly remarkable for $d$- and $f$-phase shifts. For example, the  $\delta_{d}^{\rm Ar@C_{60}}$, as a function of $\epsilon$, develops a broad minimum
with decreasing $\epsilon$, whereas the corresponding minimum in $\delta_{d}^{\rm Xe@C_{60}}$ is noticeably narrower than in $\delta_{d}^{\rm Ar@C_{60}}$. In contrast,
as one moves from the compact Ar and Xe atoms to a soft Ba atom, one sees the emergence of the well-developed low-energy narrow maximum in
$\delta_{d}^{\rm Ba@C_{60}}$ rather than a minimum.  Talking about the $f$-phase shifts, one notices how a practically
monotonic behavior of $\delta_{f}^{\rm Ar@C_{60}}$ versus $\epsilon$ is replaced by the presence of  the broad minimum in   $\delta_{f}^{\rm Xe@C_{60}}$. The latter, in turn,
changes to the much narrower minimum in $\delta_{f}^{\rm Ba@C_{60}}$ shifted toward lower electron energies. This is illustrative of the sensitivity of the electron elastic-scattering phase shifts to the size and softness (or compactness) of an encapsulated atom.

The authors will renew the discussion of the $\delta_{\ell}^{A@{\rm C}_{60}}$ phase shifts with $\ell \le 3$  later in this paper. At present, in order to continue the in-breadth discussion
of the phase-shift topic, let us explore
calculated data for $\delta_{\ell}^{A@{\rm C}_{60}}$s and $\delta_{\ell}^{\rm C_{60}}$s with $\ell \ge 4$,  Fig.~\ref{FigArXeBa49}.

Note that there are certain energy regions where  $\delta_{\ell \ge 4}^{\rm A@C_{60}}$s and $\delta_{\ell \ge 4}^{\rm C_{60}}$s are indistinguishable from each other. This implies that an incident electron with $\ell \ge 4$ stops ``detecting'' the presence of an encapsulated atom inside C$_{60}$, at given energies. For example, for $\ell=4$, the phase shifts upon electron elastic scattering off any of the considered fullerenes are the same in the energy region of approximately $0$ to $2.5$ eV. At a greater energy, the graph for
 $\delta_{\ell=4}^{\rm Ba@C_{60}}$
splits off the rest of the graphs. The splitting increases with increasing energy. In contrast, a close equivalency between $\delta_{\ell = 4}^{\rm Ar@C_{60}}$,
 $\delta_{\ell = 4}^{\rm Xe@C_{60}}$, and $\delta_{\ell = 4}^{\rm C_{60}}$ remains in the energy region up to $\epsilon \approx 15$ eV. Beyond this energy, it is the graph for
 $\delta_{\ell = 4}^{\rm Xe@C_{60}}$ which splits off the graphs for $\delta_{\ell = 4}^{\rm Ar@C_{60}}$ and  $\delta_{\ell = 4}^{\rm C_{60}}$, and splitting increases with increasing energy.
 In turn, the phase shifts $\delta_{\ell = 4}^{\rm Ar@C_{60}}$ and  $\delta_{\ell = 4}^{\rm C_{60}}$ remain equal to each other in a broader energy region, up to $\epsilon \approx 25$ eV.
For higher orbital quantum numbers,
$\ell > 4$, the energy region, where there is the equivalency between all phase shifts of interest expands noticeably with increasing $\ell$. For example, for compact
Ar@C$_{60}$ and Xe@C$_{60}$ as well as empty C$_{60}$, corresponding phase shifts become and remain identical in the whole considered energy domain when the orbital quantum number reaches the value of $\ell=9$. For soft Ba@C$_{60}$, however, the situation is clearly different, and  $\delta_{\ell \ge 4}^{\rm Ba@C_{60}}$ remains noticeably split off the rest of the graphs even at $\ell=9$. Once again, the discussed  results uncover  the sensitivity of electron elastic scattering by $A$@C$_{60}$ to the size, compactness, and softness of the encapsulated atom.

The authors now return to the previously postponed discussion of phase shifts  $\delta_{\ell \le 3}^{A@{\rm C}_{60}}$ and $\delta_{\ell \le 3}^{\rm C_{60}}$ for orbital quantum numbers of an incident electron $\ell \le 3$ (see Fig.~\ref{FigArXeBa03}) on a more detailed scale.

First, let us focus the reader's attention on the phase shift values at $\epsilon =0$, see Table~\ref{Table1}.
\begin{table}
\caption{\label{Table1} Calculated HF electron elastic-scattering phase shifts $\delta_{\ell}(\epsilon)$ (at $\epsilon =0$) upon electron collision with empty C$_{60}$ and $A$@C$_{60}$ ($A={\rm Ar, Xe}$, and ${\rm Ba}$).}
\begin{ruledtabular}
\begin{tabular}{ddddd}
\ell &    \multicolumn{4}{c}{$\delta_{\ell}(0)$}   \\ \cline{2-5}
      & {\rm C_{60}}  & {\rm Ar}  &  {\rm Xe} &  {\rm Ba} \\
 \hline
s & \pi & 4\pi & 6\pi & 6\pi \\
p & \pi & 3\pi & 5\pi & 5\pi \\
d & \pi & \pi  & 3\pi & 3\pi \\
f &  0  & 0    & 0    & \pi  \\
\end{tabular}
\end{ruledtabular}
\end{table}

In order to understand the behavior of phase shifts at $\epsilon~\rightarrow~0$, let us refer to Levinson theorem \citep{Landau} which  we write as follows:
\begin{eqnarray}
\left. \delta_{\ell}(\epsilon)\right\vert_{\epsilon \rightarrow 0}\rightarrow (N_{n_{\ell} }+ q_{\ell })\pi.
\label{Levinson}
\end{eqnarray}
Here, $N_{n_{\ell}}$ is the number of occupied states with given $\ell$ in
the ground-state configuration of a target-scatterer, whereas $q_{\ell }$ is the
number of additional (if any) empty bound states with the same $\ell$ which can accommodate (bind) an external electron.
For the \textit{empty} C$_{60}$ cage approximated by the annular potential, Eq.~(\ref{SWP}), $N_{n_{\ell}}=0$ for all $\ell$s. Therefore,
from the calculated values of $\delta_{\ell}^{\rm C_{60}}(0)$, Table~\ref{Table1}, one concludes that $q_{\ell }=1$ for $\ell = s, p$, and $d$,
but $q_{\ell}=0$ for $\ell = f$. The implication is that the confining potential $U_{\rm c}(r)$ (or the C$_{60}$ cage itself) has the ability
to bind an electron into a $s$-, or $p$-, or $d$-state; this was already noted in Ref.~\citep{Winstead}. In addition to results of Ref.\cite{Winstead},
the present study predicts the existence of the $s$-, $p$-, and $d$-anions Ar@C$_{60}^{-}$ and Xe@C$_{60}^{-}$, in the given approximation. Indeed, if one counts the number of occupied $s$-, $p$-,
and $d$-subshells in Ar ($N_{n_{\ell}}= 3, 2$, and $0$, respectively) and Xe ($N_{n_{\ell}}= 5, 4$, and $2$, respectively), then, with the help of Eq.~(\ref{Levinson}) and Table~\ref{Table1}, one
easily finds
that $q_{s}=q_{p}=q_{d} =1$ whereas $q_{f}=0$. For Ba@C$_{60}$, however, the situation is somewhat different. Indeed,
as shown in Table~\ref{Table1}, $\delta_{s}^{\rm Ba@C_{60}}(0) = 6\pi$, and there are exactly six  $s$-subshells in the Ba atom, i.e., $N_{n_{s}}=6$. This makes  $q_{s}=0$, for Ba@C$_{60}$. The
latter  indicates the absence of
a $s$-anion Ba@C$_{60}^{-}$, in contrast to the situation for Ar@C$_{60}$, Xe@C$_{60}$, and C$_{60}$. Next, note that $\delta_{f}^{\rm Ba@C_{60}}(0) = \pi$, although there are no occupied $f$-subshells in the Ba atom.
Hence, $q_{f}=1$. This predicts the existence of a $f$-anion Ba@C$_{60}^{-}$, again in contrast to the case of the Ar@C$_{60}$, Xe@C$_{60}$, and C$_{60}$ fullerenes.

Furthermore, note that the plotted phase shifts have oscillatory structures throughout the whole energy region. For empty C$_{60}$, this is due to interference
between the incident electronic wave and electronic wave scattered off the C$_{60}$ cage. For $A$@C$_{60}$, this  is because of interference between the incident electronic wave and the
electronic waves scattered off the C$_{60}$ cage and off the encaged atom $A$ of the fullerene.

What catches one's eye, though, is the well-developed low-energy \textit{maximum}
in the $d$-phase shift $\delta_{d}^{\rm Ba@C_{60}}$ in contrast to the low-energy \textit{minima} in the $d$-phase shifts $\delta_{d}^{\rm Xe@C_{60}}$ and $\delta_{d}^{\rm Ar@C_{60}}$.
Additionally, the noted minimum in $\delta_{d}^{\rm Ar@C_{60}}$ is much broader than the minimum in $\delta_{d}^{\rm Xe@C_{60}}$. Next, there is the well-developed low-energy minimum in
the $f$-phase shifts $\delta_{f}^{\rm Ba@C_{60}}$, but much broader and shallower minimum in $\delta_{f}^{\rm Xe@C_{60}}$, and no such minimum emerges in $\delta_{f}^{\rm Ar@C_{60}}$.
It is not at all clear why the phase shifts behave like that. In the following, the authors elucidate the reason for the behavior of the phase shifts in question.

\begin{figure}[ht]
\center{\includegraphics[width=8cm]{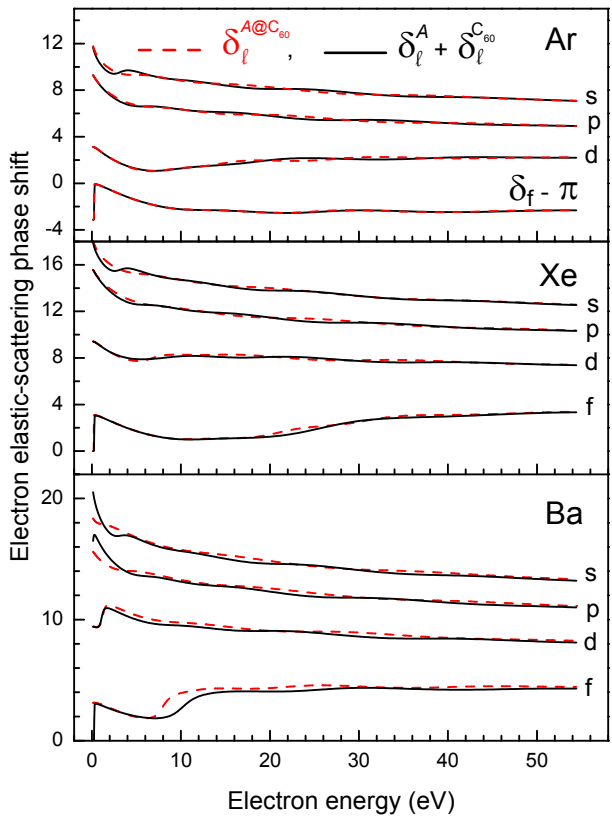}}
\caption{Calculated HF phase shifts (in units of radian) $\delta_{\ell}^{A \rm @C_{60}}$, as well as ${\tilde \delta}_{\ell}^{A \rm @C_{60}}$, Eq.~(\ref{EqD+D}),  upon electron elastic scattering off
Ar@C$_{60}$, Xe@C$_{60}$, and Ba@C$_{60}$ as marked.}
\label{FigDD}
\end{figure}

It is found in the present study that, as odious as it may seem, the above observations can be understood in terms of a simple sum of a phase shift  $\delta_{\ell}^{\rm C_{60}}$ due to
electron scattering off empty C$_{60}$ and a phase shift $\delta_{\ell}^{A}$ upon electron scattering by the \textit{isolated}
atom $A$ (recently, the other authors \citep{Amusia15} have come to the same conclusion as well). The above stated approximation will be referred to as
the \textit{independent-scattering approximation} in the present paper.
 Correspondingly,
\begin{eqnarray}
\delta_{\ell}^{A \rm @C_{60}}(\epsilon) \approx {\tilde \delta}_{\ell}^{A \rm @C_{60}}(\epsilon) = \delta_{\ell}^{A}(\epsilon) + \delta_{\ell}^{\rm C_{60}}(\epsilon).
\label{EqD+D}
\end{eqnarray}
The qualitative usability of Eq.~(\ref{EqD+D}) is evident from the comparison between plotted in Fig.~\ref{FigDD} $\delta_{\ell}^{A \rm @C_{60}}$ (dashed line) and ${\tilde \delta}_{\ell}^{A \rm @C_{60}}$ (solid line).

One can see from Fig.~\ref{FigDD} that, indeed ${\tilde \delta}_{\ell}^{A \rm @C_{60}} \approx \delta_{\ell}^{A \rm @C_{60}}$, to a good approximation, although not without exceptions.
 The exceptions are mostly noticeable for the $s$- and $p$-phase shifts for electron scattering off Ba@C$_{60}$ below $5$ eV. The Ba atom, however, is a less suitable atom
 to apply the independent-scattering approximation to.
This is because Ba, in contrast to Ar and Xe,  transfers much of its valence electron density to the C$_{60}$ cage.
 Anyway, one, of course, would be too naive to expect that the independent-scattering approximation is anywhere perfect. The usability of it, as one can see,
is somewhat limited.

\begin{figure}[ht]
\center{\includegraphics[width=6cm]{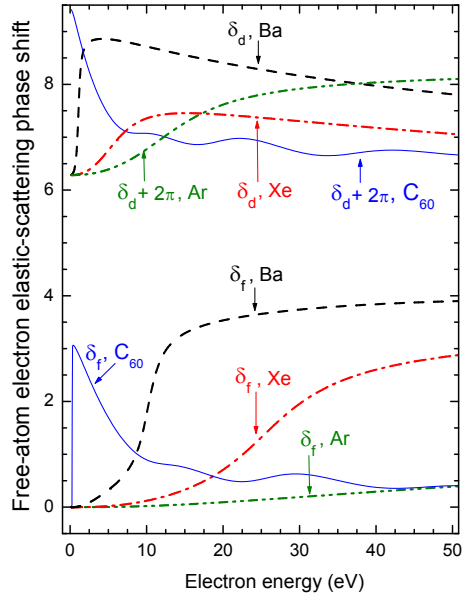}}
\caption{Calculated HF free-atom phase shifts (in units of radians) $\delta_{\ell}^{\rm Ar}$, $\delta_{\ell}^{\rm Xe}$, $\delta_{\ell}^{\rm Ba}$, and $\delta_{\ell}^{\rm C_{60}}$, as marked.}
\label{Fig5}
\end{figure}

Let us demonstrate how   Eq.~(\ref{EqD+D}) helps one to understand the emergence of the well-developed low-energy maximum in $\delta_{d}^{\rm Ba@C_{60}}$ but the minimum in $\delta_{d}^{\rm Xe@C_{60}}$.
For this, let us explore Fig.~\ref{Fig5} where plotted are the calculated HF free-atom $d$-phase shifts $\delta_{d}^{\rm Xe}$ and $\delta_{d}^{\rm Ba}$ along with
 $\delta_{d}^{\rm C_{60}}$.

One can see from  Fig.~\ref{Fig5} that, as $\epsilon$ decreases down to approximately $10$ eV, both $\delta_{d}^{\rm Xe}$ and $\delta_{d}^{\rm C_{60}}$ increase slowly at about an equal rate. Thus, the $\delta_{d}^{\rm Xe@C_{60}}$ phase shift should increase slowly with decreasing $\epsilon$ as well, according to Eq.~(\ref{EqD+D}). At lower energies, between $\epsilon \approx 10$ and $5$ eV, the free-Xe
$\delta_{d}^{\rm Xe}$ phase shift decreases at a greater rate than the increasing
phase shift $\delta_{d}^{\rm C_{60}}$. Therefore, their sum  starts decreasing in this energy region, and so should do $\delta_{d}^{\rm Xe@C_{60}}$ as well. At yet lower energies, below $\epsilon \approx 5$ eV, however, the increase of
$\delta_{d}^{\rm C_{60}}$ outpaces strongly the decrease of free-Xe $\delta_{d}^{\rm Xe}$. The sum  of these two phase shifts starts increasing with decreasing $\epsilon$ in this energy region, and so should do
$\delta_{d}^{\rm Xe@C_{60}}$ as well. All of the above results in the emergence of the \textit{minimum} in $\delta_{d}^{\rm Xe@C_{60}}$ below $\epsilon \approx 10$ eV.
A close agreement between the thus predicted results for $\delta_{d}^{\rm Xe@C_{60}}$ and calculated results obtained on the basis of the endohedral HF Eq.~(\ref{eqHF})
 was demonstrated in Fig.~\ref{FigDD}.

 A similar analysis of the sum of the free-atom $d$-phase shift $\delta_{d}^{\rm Ba}$ and $\delta_{d}^{\rm C_{60}}$ (see Fig.~\ref{Fig5}) can easily lead one to the prediction
 of a low-energy \textit{maximum} in $\delta_{d}^{\rm Ba@C_{60}}$ below $10$ eV. A close agreement between the thus predicted results for $\delta_{d}^{\rm Ba@C_{60}}$ and calculated results obtained on the basis of the endohedral HF Eq.~(\ref{eqHF}) was demonstrated in Fig.~\ref{FigDD} as well.

It now becomes clear that the emergence of the minimum in $\delta_{d}^{\rm Xe@C_{60}}$ but maximum
in $\delta_{d}^{\rm Ba@C_{60}}$, below $\epsilon \approx 10$ eV, is due to two reasons. First, this is because the free-Xe  $\delta_{d}^{\rm Xe}$ phase shift starts falling down at a higher electron energy
($\epsilon \approx 11$ eV) than the free-Ba $\delta_{d}^{\rm Ba}$ phase shift ($\epsilon \approx 2.5$ eV). Second, this is because the decrease of $\delta_{d}^{\rm Ba}$ occurs at a much greater rate than the decrease of $\delta_{d}^{\rm Xe}$ in the corresponding electron energy region. Thus, the discussed marked differences between $\delta_{d}^{\rm Xe@C_{60}}$ and $\delta_{d}^{\rm Ba@C_{60}}$ are associated primarily with the individuality
of the free-Xe and free-Ba $d$-phase shifts, at a given phase shift $\delta_{d}^{\rm C_{60}}$.

It also becomes clear why the low-energy minimum in $\delta_{d}^{\rm Ar@C_{60}}$ is broader than that in the above discussed $\delta_{d}^{\rm Xe@C_{60}}$ (see Fig.~\ref{FigArXeBa03}).
This is because (see Fig.~\ref{Fig5}) the decrease of free-Ar $\delta_{d}^{\rm Ar}$ with decreasing energy
occurs at much higher energies and at a noticeably slower rate than the decrease of free-Xe $\delta_{d}^{\rm Xe}$ in the whole energy region. As a result,
this leads to the broader and shallow minimum in $\delta_{d}^{\rm Ar@C_{60}}$ than in $\delta_{d}^{\rm Xe@C_{60}}$.

It also becomes clear why  there is the well-developed low-energy minimum in
the $f$-phase shift $\delta_{f}^{\rm Ba@C_{60}}$, the much broader and shallower minimum in $\delta_{f}^{\rm Xe@C_{60}}$, but no minimum in $\delta_{f}^{\rm Ar@C_{60}}$
(see Fig.~\ref{FigArXeBa03}). This is because (see Fig.~\ref{Fig5}) the decrease of free-Ba $\delta_{f}^{\rm Ba}$ with decreasing energy is sharper and occurs at the lower energy than the decrease
of free-Xe $\delta_{f}^{\rm Xe}$. This decrease of $\delta_{f}^{\rm Xe}$, in turn, is sharper and occurs at a lower energy than the decrease
of free-Ar $\delta_{f}^{\rm Ar}$. This induces the above noted differences
between $\delta_{f}^{\rm Ba@C_{60}}$, $\delta_{f}^{\rm Xe@C_{60}}$, and $\delta_{f}^{\rm Ar@C_{60}}$.
\subsubsection{Electron elastic-scattering and bremsstrahlung cross sections}
\begin{figure*}[ht]
\center{\includegraphics[width=16cm]{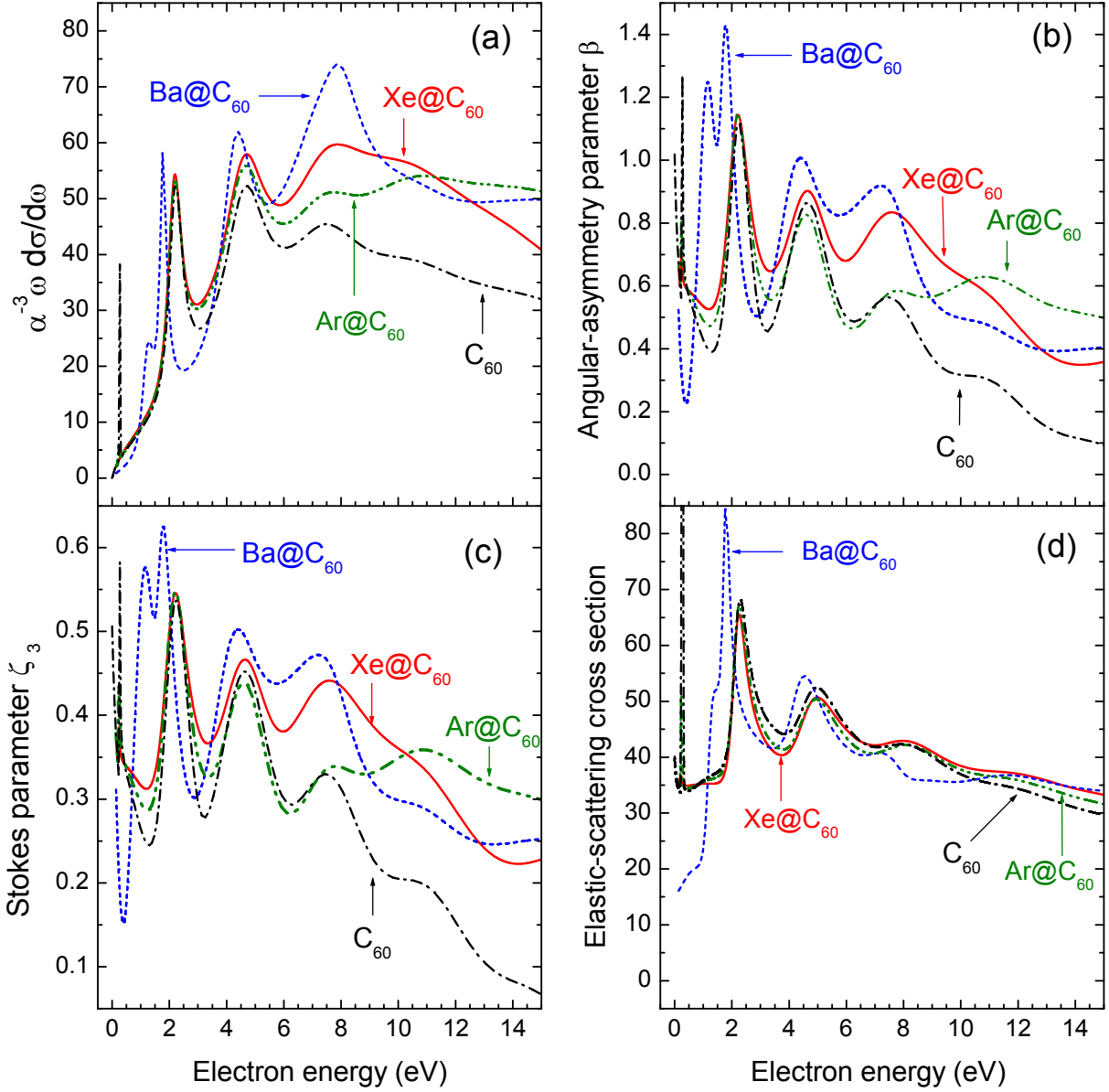}}
\caption{Calculated HF $\alpha^{-3} \omega\frac{d\sigma}{d\omega}$ (in atomic units), angular asymmetry parameter $\beta(\epsilon)$, and Stokes polarization parameter $\zeta_{3}(\epsilon)$ at $\theta=90^{\circ}$ of electron low-frequency bremsstrahlung
on $A$@C$_{60}$ ($A= \rm Ar, Xe$, and Ba) and empty C$_{60}$, as well as corresponding total
electron elastic-scattering cross sections $\sigma_{\rm el}^{A@\rm C_{60}}$ and $\sigma_{\rm el}^{\rm C_{60}}$ (in units of $20\,a_{0}^2$), as marked.}
\label{FigArXeBa}
\end{figure*}

Calculated spectral density $d\sigma/d\omega$, angular-asymmetry parameter $\beta(\epsilon)$, and Stokes polarization-parameter $\zeta_{3}(\epsilon)$$|_{\theta=90^{\circ}}$ of low-frequency bremsstrahlung as well as total electron elastic-scattering cross sections $\sigma_{\rm el}^{A@\rm C_{60}}$ due to electron collision with Ar@C$_{60}$, Xe@C$_{60}$, Ba@C$_{60}$, and empty C$_{60}$, are depicted in Fig.~\ref{FigArXeBa}.

 One learns from Fig.~\ref{FigArXeBa} that the bremsstrahlung parameters are strongly oscillating functions of electron energy $\epsilon$ in the considered region of $\epsilon \le 15$ eV. Also oscillatory functions of energy are the total electron elastic-scattering cross sections; the latter is in accordance with results of earlier works \citep{e+A@C60,Winstead}. What catches one's eye, though, is the following. First, the spectral density
 $d\sigma^{A@{\rm C_{60}}}/d\omega$ of bremsstrahlung behaves quite differently than the total electron elastic-scattering cross section $\sigma_{\rm el}^{A@\rm C_{60}}$, especially for the case of Ba@C$_{60}$. Indeed, whereas the intensities of the first three maxima in each of $\sigma_{\rm el}^{A@\rm C_{60}}$ are weakening with increasing energy, the intensities of the corresponding maxima in  $d\sigma^{A@{\rm C_{60}}}/d\omega$ are, generally, not.
Moreover, surprisingly, differences in the depicted bremsstrahlung parameters between the four systems - Ar@C$_{60}$, Xe@C$_{60}$, Ba@C$_{60}$, and empty C$_{60}$ - are much more
prominent than the differences between the corresponding total electron elastic-scattering cross sections $\sigma_{\rm el}$. Indeed, e.g., there are practically no differences exist between
$\sigma_{\rm el}^{\rm Ar@C_{60}}$ and $\sigma_{\rm el}^{\rm Xe@C_{60}}$, and both of them differ insignificantly from $\sigma_{\rm el}^{\rm C_{60}}$ in the whole energy region. This is indicative of a little impact of the presence of a ``compact'' encapsulated atom $A$ on ${\rm e}+A@\rm{C_{60}}$ elastic scattering. On the contrary, noticeable, if not significant,
differences in the bremsstrahlung parameters between all four systems are indisputable. The described results, thus, uncover that electron bremsstrahlung is more sensitive to the presence of a particular atom inside of C$_{60}$  than the corresponding total electron elastic scattering cross section. This is valid even if the encapsulated atom is a ``compact'' atom. The authors attribute this primarily to that fact that the spectral density of bremsstrahlung depends on differences between electron elastic-scattering phase shifts, whereas $\sigma_{\rm el}^{A@\rm C_{60}}$ depends on the absolute values of the phase shifts.

\subsection{Electron collision with high-spin $A$@C$_{60}$: A = N, Cr, Mn, Mo, Tc, and Eu}

\subsubsection{N@C$_{60}$}

\begin{figure}[ht]
\center{\includegraphics[width=8cm]{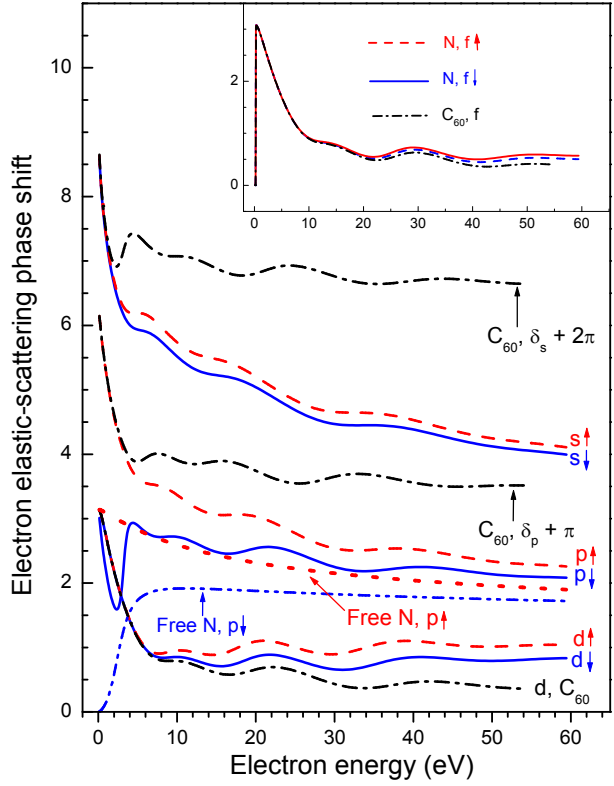}}
\caption{Calculated SPHF $s$-, $p$-, $d$-, and $f$-phase shifts (in units of radian), as marked. Solid lines, $\delta_{\ell\downarrow}^{\rm N@C_{60}}$s due to scattering of spin-down electrons off
N@C$_{60}$. Dashed lines, $\delta_{\ell\uparrow}^{\rm N@C_{60}}$s due to scattering of spin-up electrons off N@C$_{60}$. Dash-dotted lines, $\delta_{\ell}^{\rm C_{60}}$s due to
electron scattering off empty C$_{60}$ (these phase shifts are spin-independent in the utilized approximation). Dash-dot-dotted and dotted lines, the $p$-phase shifts due to electron scattering off spin-up and spin-down electrons, respectively, off free N.}
\label{FigN03}
\end{figure}

The nitrogen atom
N(${1\rm s^{1}}$$\uparrow$${1\rm s^{1}}$$\downarrow$${2\rm s^{1}}$$\uparrow$${2\rm s^{1}}$$\downarrow$$\rm 2p^{3}$$\uparrow$,\,$^{4}$S) is the first atom with a multielectron semifilled subshell (the $2p^{3}$ subshell) in the periodic table, thus being the smallest one among other high-spin  multielectron atoms. N@C$_{60}$, therefore, serves as a good starting sample for the discussion of electron collision with high-spin endohedral fullerenes. Corresponding calculated SPHF  electron elastic-scattering phase shifts are depicted in Fig.~\ref{FigN03}.

Note how elastic-scattering phase shifts of incident spin-up electrons, $\delta_{\ell\uparrow}^{\rm N@C_{60}}$, differ from those $\delta_{\ell\downarrow}^{\rm N@C_{60}}$ of spin-down electrons.
This can be easily understood in the framework of the SPHF theory. Namely, the differences  are primarily due to the presence (absence) of exchange interaction between incident spin-up (spin-down)
electrons with the electrons of the spin-unpaired spin-up $2p^{3}$$\uparrow$-subshell of the atom.

Furthermore, note particularly dramatic differences between the phase shifts $\delta_{p\downarrow}^{\rm N@C_{60}}$ and
$\delta_{p\uparrow}^{\rm N@C_{60}}$ of the $p$$\downarrow$- and $p$$\uparrow$-scattered electronic waves. These  phase shifts are seen to be taking  different routes with decreasing $\epsilon$.
 Indeed, the $\delta_{p\uparrow}^{\rm N@C_{60}}$ phase shift is, on the average, monotonically increasing to $\delta_{p\uparrow}^{\rm N@C_{60}}(0) = 2\pi$
with decreasing $\epsilon$.
In contrast, the $\delta_{p\downarrow}^{\rm N@C_{60}}$ phase shift has a deep narrow minimum at $\epsilon \approx 2$ eV, after which $\delta_{p\downarrow}^{\rm N@C_{60}} \rightarrow \pi$ at $\epsilon \rightarrow 0$.

In order to unveil the origin of the dramatic difference between  $\delta_{p\downarrow}^{\rm N@C_{60}}$ and $\delta_{p\uparrow}^{\rm N@C_{60}}$, let us exploit both Levinson theorem, Eq.~(\ref{Levinson}), and the 
independent-scattering approximation, Eq.~(\ref{EqD+D}).
In accordance with the latter,
$\delta_{p\downarrow(\uparrow)}^{\rm N@C_{60}}(\epsilon) \approx \delta_{p\downarrow(\uparrow)}^{\rm N}(\epsilon) + \delta_{p}^{\rm C_{60}}(\epsilon)$, where $\delta_{p\downarrow(\uparrow)}^{\rm N}$ is the
elastic-scattering phase shift of a spin-down(spin-up) electron upon its collision with a free N atom. The free-N phase shifts $\delta_{p\downarrow}^{\rm N}$ and $\delta_{p\uparrow}^{\rm N}$
 are depicted in Fig.~\ref{FigN03} by the dash-dot-dotted and dotted lines, respectively.

Note how $\delta_{p\downarrow}^{\rm N}$ starts, at $\epsilon \approx 7$ eV, abruptly dropping to a zero at $\epsilon \rightarrow 0$. This is in accordance with Levinson theorem, because there are no $np$$\downarrow$ bound-states
either in the  ground-state structure of free N or in the field of N, in the SPHF approximation. It is now easy to see that the sum of the two phase shifts $\delta_{p\downarrow}^{\rm N}$ and $\delta_{p}^{\rm C_{60}}$
translates into the originally predicted low-energy minimum  in $\delta_{p\downarrow}^{\rm N@C_{60}}$.

Next, note that, in contrast to $\delta_{p\downarrow}^{\rm N}$, the phase shift  $\delta_{p\uparrow}^{\rm N} \rightarrow \pi$ at $\epsilon \rightarrow 0$. This, again, is in line with Levinson theorem, because there is only one spin-unpaired spin-up subshell with $\ell=1$
(the $2p^{3}$$\uparrow$ subshell) in the nitrogen configuration. Correspondingly, this time, the sum of monotonically increasing $\delta_{p\uparrow}^{\rm N}$ and $\delta_{p}^{\rm C_{60}}$ translates into \textit{steadily} increasing
$\delta_{p\uparrow}^{\rm N@C_{60}}$ with decreasing energy, thereby resulting in the marked differences in question between $\delta_{p\uparrow}^{\rm N@C_{60}}$ and  $\delta_{p\downarrow}^{\rm N@C_{60}}$.

Next, a trial calculation showed that, for partial electronics waves with $\ell \ge 4$, $\delta_{\ell\uparrow}^{\rm N@C_{60}} \approx \delta_{\ell\downarrow}^{\rm N@C_{60}} \approx \delta_{\ell}^{\rm C_{60}}$.
Therefore, because $\delta_{\ell\ge 4}^{\rm C_{60}}$s have already been plotted earlier in the paper, the phase shifts
$\delta_{\ell\uparrow(\downarrow)}^{\rm N@C_{60}}$ with $\ell \geq 4$ are not plotted separately in the present paper.

\begin{figure*}[ht]
\center{\includegraphics[width=16cm]{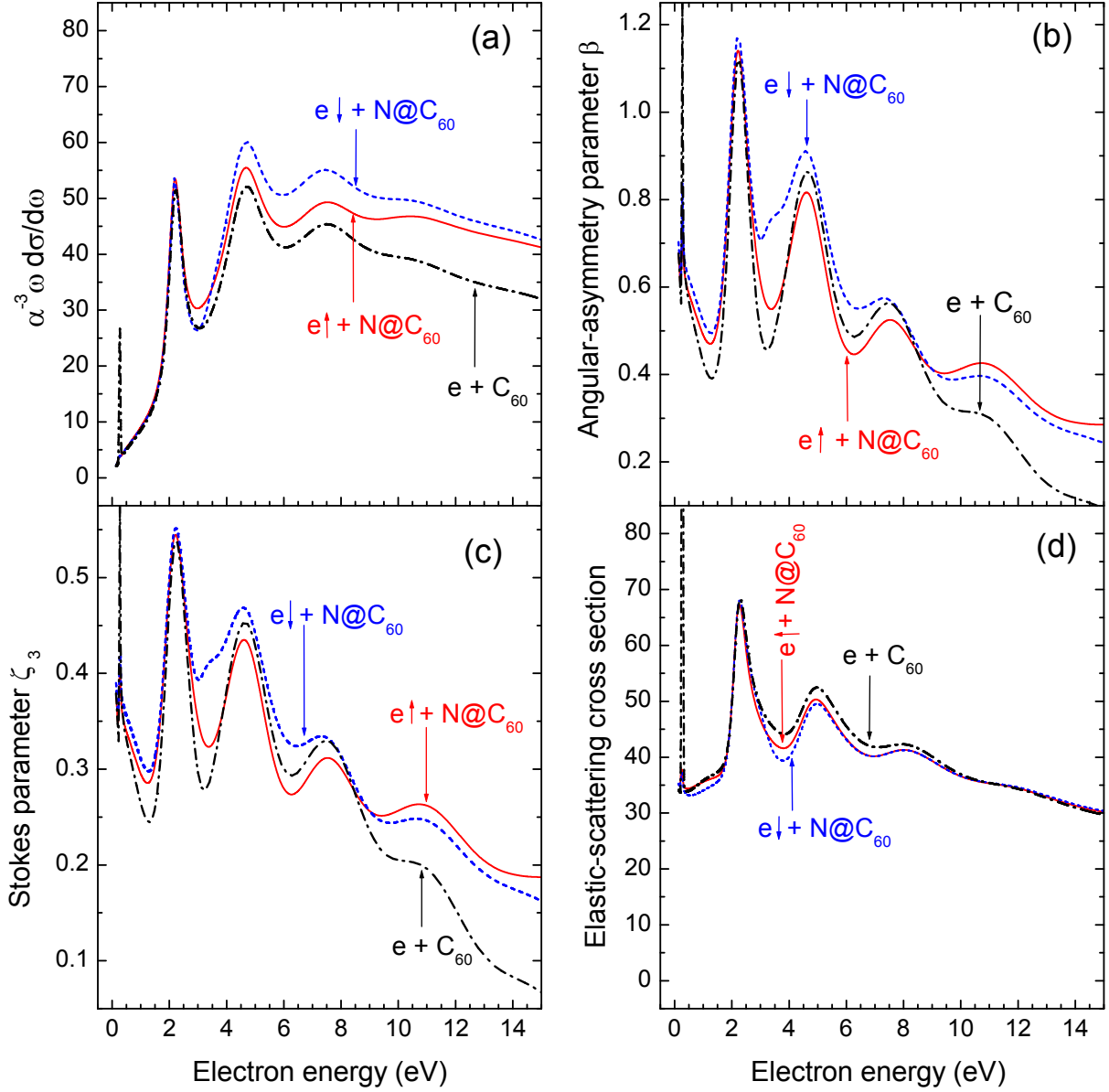}}
\caption{Calculated SPHF
$\alpha^{-3} \omega\frac{d\sigma_{\uparrow(\downarrow)}}{d\omega}$ (in atomic units), angular asymmetry parameter $\beta_{\uparrow(\downarrow)}$, and Stokes polarization parameter
$\zeta_{3}$$\uparrow$$(\downarrow)$$|_{\theta=90^{\circ}}$ of electron low-frequency bremsstrahlung,
as well as the  total
electron elastic-scattering cross section $\sigma_{\rm el\uparrow(\downarrow)}$  (in units of $20\,a_{0}^2$) upon electron collision both with N@C$_{60}$ and C$_{60}$, as marked.}
\label{FigN}
\end{figure*}

Finally, the spectral density $d\sigma_{\uparrow(\downarrow)}/d\omega$, angular-asymmetry parameter $\beta_{\uparrow(\downarrow)}$, and Stokes polarization parameter
$\zeta_{3}$$\uparrow$$(\downarrow)$$|_{\theta=90^{\circ}}$ of low-frequency bremsstrahlung as well as the corresponding total
electron elastic-scattering cross section $\sigma_{\rm el\uparrow(\downarrow)}$ due to collision of a spin-up (spin-down) electron with N@C$_{60}$ are depicted in Fig.~\ref{FigN}.

One can see from Fig.~\ref{FigN} that the bremsstrahlung parameters, as well as
$\sigma_{\rm el\uparrow(\downarrow)}$ are strongly oscillating functions of electron energy $\epsilon$, as in the case of electron scattering off Ar@C$_{60}$, Xe@C$_{60}$, and Ba@C$_{60}$.
However, most interesting is that whereas $\sigma_{\rm el\uparrow}$ and $\sigma_{\rm el\downarrow}$
practically do not differ from each other, the differences between electron spin-up and spin-down bremsstrahlung off N@C$_{60}$ are significant.
Once again we encounter the situation where electron bremsstrahlung is more sensitive, compared to the electron elastic-scattering cross section, to the presence of the atom inside C$_{60}$.

\subsubsection{Cr@C$_{60}$ and Mn@C$_{60}$}

The first two atoms in the periodic table with a more capacious spin-unpaired semifilled subshell in their ground-states than the $2p^{3}$$\uparrow$ subshell of N are
the Cr(...$3d^{5}$$\uparrow$$4s^{1}$$\uparrow$,\,$^{7}S$) and Mn(...$3d^{5}$$\uparrow$$4s^{1}$$\uparrow$$4s^{1}$$\downarrow$,\,$^{6}S$)  atoms. Both of them possess a spin-unpaired
$3d^{5}$$\uparrow$ semifilled subshell. Additionally, the Cr atom has a second semifilled subshell which is the $4s^{1}$$\uparrow$ subshell. Moreover, both
atoms are soft atoms, in contrast to N. They
donate a noticeable part of their $4s$-valence electron density to the C$_{60}$ cage, making it ``spin-charged''. The Cr atom ``charges'' the C$_{60}$ cage by a spin-up electron density, whereas Mn ``charges'' the C$_{60}$ cage
primarily by a spin-down electron density, as was discussed earlier in the present paper.   Consequently, the study of electron collision with the
Cr@C$_{60}$ and Mn@C$_{60}$ systems is an interesting case study. It allows one to learn about the modification of electron elastic scattering and bremsstrahlung versus (a) the increasing number of electrons in a semifilled subshell of an encapsulated atom, (b) the increasing number of semifilled subshells in the atom, and (c) the ability of the atom to ``spin charge'' the C$_{60}$ cage either by a
spin-up, or spin-down electron density.

\begin{figure}[ht]
\center{\includegraphics[width=8cm]{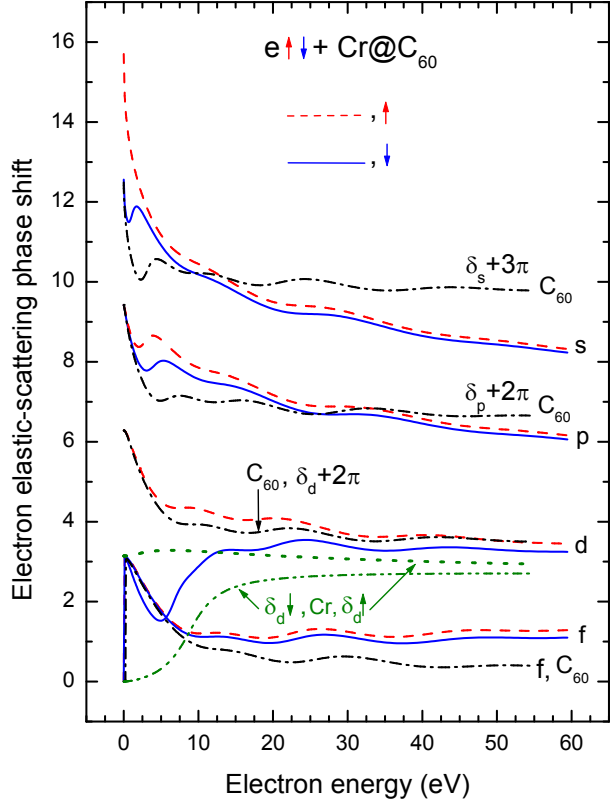}}
\caption{Calculated SPHF $s$-, $p$-, $d$-, and $f$-phase shifts (in units of radian), as marked. Solid lines, $\delta_{\ell\downarrow}^{\rm Cr@C_{60}}$s due to scattering of spin-down electrons off
Cr@C$_{60}$. Dashed lines, $\delta_{\ell\uparrow}^{\rm Cr@C_{60}}$s  due to scattering of spin-up electrons off Cr@C$_{60}$. Dash-dotted lines, $\delta_{\ell}^{\rm C_{60}}$s due to
electron scattering off empty C$_{60}$. Dash-dot-dotted and dotted lines, the free-atom $d$-phase shifts upon electron scattering of spin-down and spin-up electrons, respectively, off free Cr.}
\label{FigCr03}
\end{figure}

\begin{figure}[ht]
\center{\includegraphics[width=8cm]{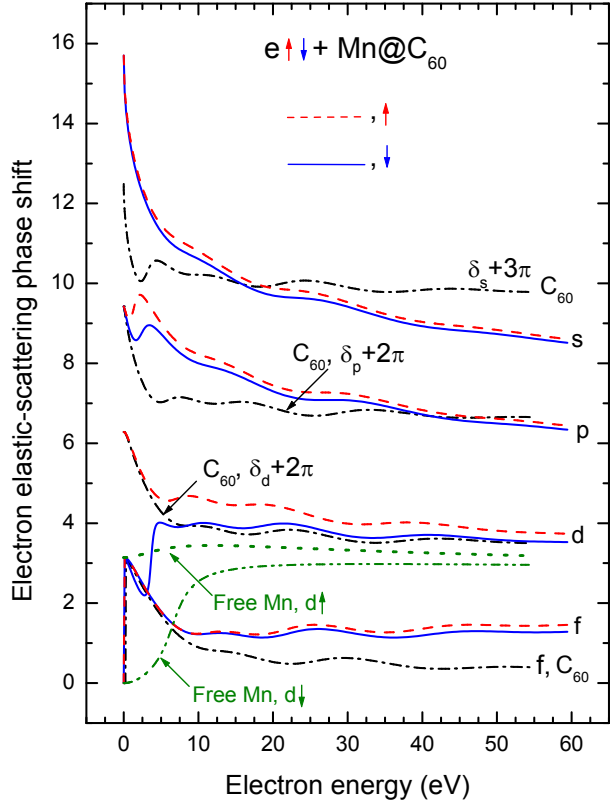}}
\caption{Calculated SPHF $s$-, $p$-, $d$-, and $f$-phase shifts (in units of radian), as marked. Solid lines, $\delta_{\ell\downarrow}^{\rm Mn@C_{60}}$s due to scattering of spin-down electrons off
Mn@C$_{60}$. Dashed lines, $\delta_{\ell\uparrow}^{\rm Mn@C_{60}}$s  due to scattering of spin-up electrons off Mn@C$_{60}$. Dash-dotted lines, $\delta_{\ell}^{\rm C_{60}}$s due to
electron scattering off empty C$_{60}$. Dash-dot-dotted and dotted lines, the free-atom $d$-phase shifts upon electron scattering of spin-down and spin-up electrons, respectively, off free Mn.}
\label{FigMn03}
\end{figure}

SPHF calculated electron elastic-scattering phase shifts of spin-up and spin-down electrons with $\ell \le 3$ due to collision with Cr@C$_{60}$ or Mn@C$_{60}$ are depicted in Figs.~\ref{FigCr03} and \ref{FigMn03},
respectively.

First, note the values of $s$-, $p$-, and $d$-phase shifts at $\epsilon=0$. These values, in conjunction with the SPHF ground-state configurations of Cr and Mn as well as Levinson theorem, speak to that
fact that Cr@C$_{60}$ and Mn@C$_{60}$ are capable of making negative ions. Namely, they can bind
either an external spin-up or spin-down electron into a $s$-, or $p$-, or $d$-state, but not into a state with $\ell \ge 3$. The binding properties of
Cr@C$_{60}$ and Mn@C$_{60}$ are, thus, the same as the binding properties of compact Ar@C$_{60}$, Xe@C$_{60}$, and N@C$_{60}$. This is interesting, because the $4s$$\uparrow$$(\downarrow)$-valence electron
density of Cr and Mn is subject to a strong drain from the encapsulated atom to the C$_{60}$ cage, in contrast the case of encapsulated compact Ar, Xe, and N (see
Fig.~\ref{FigWF}), and yet binding properties of Cr@C$_{60}$ and Mn@C$_{60}$ are the same as in the case of Ar@C$_{60}$, Xe@C$_{60}$, and N@C$_{60}$. Next, also interesting is that although the drain of the valence electron density
from encapsulated Cr and Mn to the C$_{60}$ cage is about as strong as
in the case of encapsulated Ba, the binding properties of
Cr@C$_{60}$ and Mn@C$_{60}$, nevertheless, differ from those of Ba@C$_{60}$. Indeed, the Cr@C$_{60}$ and Mn@C$_{60}$ fullerenes do bind an extra electron into a $s$-state but
cannot bind it into a $f$-state, in contrast to Ba@C$_{60}$.

Second, note how differently the $s$-phase shift $\delta_{s\downarrow}^{\rm Cr@C_{60}}$ behaves  compared to  $\delta_{s\uparrow}^{\rm Cr@C_{60}}$,
$\delta_{s\uparrow}^{\rm Mn@C_{60}}$ and $\delta_{s\downarrow}^{\rm Mn@C_{60}}$ at low energies. Namely,  $\delta_{s\downarrow}^{\rm Cr@C_{60}}$ has a well-developed oscillation near threshold, whereas
 $\delta_{s\uparrow}^{\rm Cr@C_{60}}$,
$\delta_{s\uparrow}^{\rm Mn@C_{60}}$, and $\delta_{s\downarrow}^{\rm Mn@C_{60}}$ have not. The oscillation in $\delta_{s\downarrow}^{\rm Cr@C_{60}}$
looks similar to the oscillation emerging in the $s$-phase shift $\delta_{s}^{\rm C_{60}}$ at about the same energy.
Therefore, with the independent-scattering approximation in mind, one might think that the oscillation in $\delta_{s\downarrow}^{\rm Cr@C_{60}}$ is due to the corresponding oscillation in  $\delta_{s}^{\rm C_{60}}$.
This, however, is incorrect, otherwise all other $s$-phase shifts, namely, $\delta_{s\uparrow}^{\rm Cr@C_{60}}$, $\delta_{s\uparrow}^{\rm Mn@C_{60}}$, and $\delta_{s\downarrow}^{\rm Mn@C_{60}}$
would have possessed the same oscillation as well. Thus, this is the case where the independent-scattering approximation fails to provide the answer to the question. The authors believe that a reason for the oscillation
in $\delta_{s\downarrow}^{\rm Cr@C_{60}}$ but its absence in the other $s$-phase shifts under discussion is due to the following. Let us recall that the Cr has the spin-unpaired spin-up subshell $4s$$\uparrow$.
This induces an additional difference between exchange interactions of incident spin-up and
spin-down $s$-electrons with the electrons of the Cr atom. Perhaps, this difference, in turn, either   facilitates (for the $s$$\downarrow$-electronic wave) or cancels
(for the $s$$\uparrow$-electronic wave) a three-wave interference
between the corresponding incident $s$-electronic wave and electronic waves scattered off C$_{60}$ and encapsulated Cr, near threshold. As a result, $\delta_{s\downarrow}^{\rm Cr@C_{60}}$ has but
$\delta_{s\uparrow}^{\rm Cr@C_{60}}$ does not have the interference maximum (or minimum) near threshold. If the proposed understanding is correct, then it becomes also clear why neither of the $s$-phase shifts
$\delta_{s\uparrow}^{\rm Mn@C_{60}}$ and $\delta_{s\downarrow}^{\rm Mn@C_{60}}$ has the oscillation near threshold. This is because the Mn has both the $4s$$\uparrow$ and $4s$$\downarrow$ electrons in its $4s^{2}$ closed subshell. Correspondingly, there is no difference between exchange interaction of incident spin-up and
spin-down $s$-electrons with the $4s$-electrons of Mn. Therefore,  a three-wave interference
between the incident $s$-electronic wave and electronic waves scattered off C$_{60}$ and encapsulated Mn is presumably canceled (as in the case of the $s$$\uparrow$-wave in $e+{\rm Cr@C_{60}}$ collision).
Correspondingly, no oscillation
emerges in $\delta_{s\uparrow}^{\rm Mn@C_{60}}$ and $\delta_{s\downarrow}^{\rm Mn@C_{60}}$.

Third, note how  differently the $d$$\downarrow$-phase shifts $\delta_{d\downarrow}^{\rm Cr@C_{60}}$ and  $\delta_{d\downarrow}^{\rm Mn@C_{60}}$ behave compared to the
$d$$\uparrow$-phase shifts $\delta_{d\uparrow}^{\rm Cr@C_{60}}$  and  $\delta_{d\uparrow}^{\rm Mn@C_{60}}$, respectively, at low electron energies.
This is directly related to the presence of the unpaired $3d^{5}$$\uparrow$ subshell in both atoms.
As a result (see Figs.~\ref{FigCr03} and  \ref{FigMn03}), the free-Cr and free-Mn $d$$\downarrow$-phase shifts  $\delta_{d\downarrow}^{\rm Cr} \rightarrow 0$ and $\delta_{d\downarrow}^{\rm Mn} \rightarrow 0$
at $\epsilon = 0$,  in accordance with Levinson theorem. In contrast, the free-Cr and free-Mn $d$$\uparrow$-phase shifts $\delta_{d\uparrow}^{\rm Cr} \rightarrow \pi$ and
$\delta_{d\uparrow}^{\rm Mn} \rightarrow \pi$ at $\epsilon \rightarrow 0$,
again, in accordance with the same Levinson theorem.  Such different behavior of  $\delta_{d\downarrow}^{\rm Cr@C_{60}}$ versus $\delta_{d\uparrow}^{\rm Cr@C_{60}}$ and
$\delta_{d\downarrow}^{\rm Mn@C_{60}}$ versus $\delta_{d\uparrow}^{\rm Mn@C_{60}}$, in conjunction with the independent-scattering approximation, Eq.~(\ref{EqD+D}), results in
the well-developed low-energy minimum only in  $\delta_{d\downarrow}^{\rm Cr@C_{60}}$ and
$\delta_{d\downarrow}^{\rm Mn@C_{60}}$.

Fourth,  note how the phase shift $\delta_{d\downarrow}^{\rm Cr@C_{60}}$ (Fig.~\ref{FigCr03}) differs from  $\delta_{d\downarrow}^{\rm Mn@C_{60}}$ (Fig.~\ref{FigMn03}). Namely,
 the low-energy minimum in $\delta_{d\downarrow}^{\rm Mn@C_{60}}$ is
narrower and emerges at lower energies than the minimum in $\delta_{d\downarrow}^{\rm Cr@C_{60}}$. This is because,  the free-Mn phase shift   $\delta_{d\downarrow}^{\rm Mn}$ (Fig.~\ref{FigMn03})
starts falling down to a zero at a lower energy and at a greater rate than $\delta_{d\downarrow}^{\rm Cr}$ (Fig.~\ref{FigCr03}).

\begin{figure}[ht]
\center{\includegraphics[width=8cm]{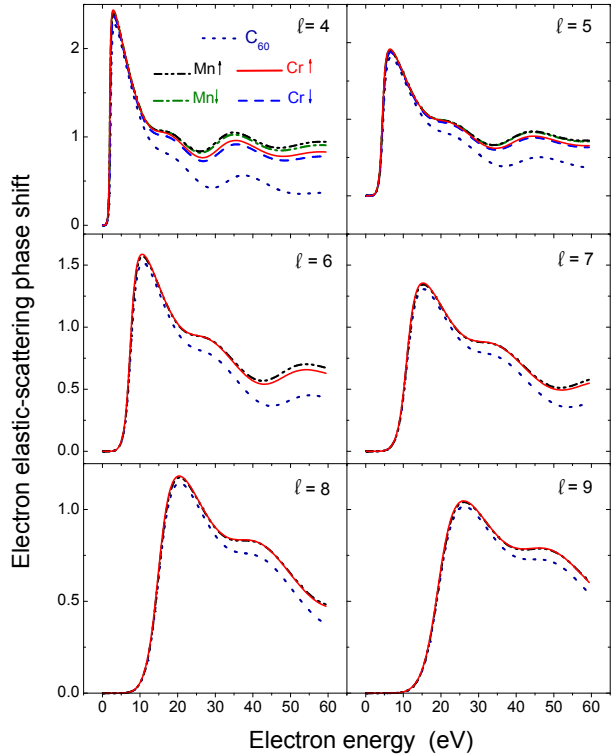}}
\caption{Calculated SPHF  electron elastic-scattering phase shifts (in units of radian) of spin-up and spin-down electrons with $\ell \ge 4$ upon their collision with Cr@C$_{60}$, Mn@C$_{60}$, and C$_{60}$, as marked.}
\label{FigCrMn49}
\end{figure}

Next, calculated SPHF electron elastic-scattering phase shifts of spin-up and spin-down electrons with $\ell \ge 4$, upon their collision with Cr@C$_{60}$ and Mn@C$_{60}$, are depicted in Fig.~\ref{FigCrMn49}.

One can see that the difference between the phase shifts of oppositely spin-polarized electrons scattered off the \textit{same fullerene} are inessential, when $\ell \ge 4$. Furthermore, note that the difference between
 the phase shifts $\delta_{\ell \ge 4}^{\rm Cr@C_{60}}$ and $\delta_{\ell \ge 4}^{\rm Mn@C_{60}}$ rapidly vanishes with increasing $\ell$ and practically disappears starting at $\ell=7$. In other words, the identity of an encapsulated atom
  inside C$_{60}$ is ``masked'' and cannot be ``resolved'' by incident electrons, when $\ell \ge 7$. Next, note that the phase shifts $\delta_{\ell \ge 4}^{\rm Cr@C_{60}}$ and
$\delta_{\ell \ge 4}^{\rm Mn@C_{60}}$ differ from the phase shifts $\delta_{\ell \ge 4}^{\rm C_{60}}$ in a broad range of energies for all $\ell$s in question. The implication is that the incoming electrons of that energy
do ``feel'' the presence of an atom inside C$_{60}$, but cannot resolve its identity.

\begin{figure*}[ht]
\center{\includegraphics[width=16cm]{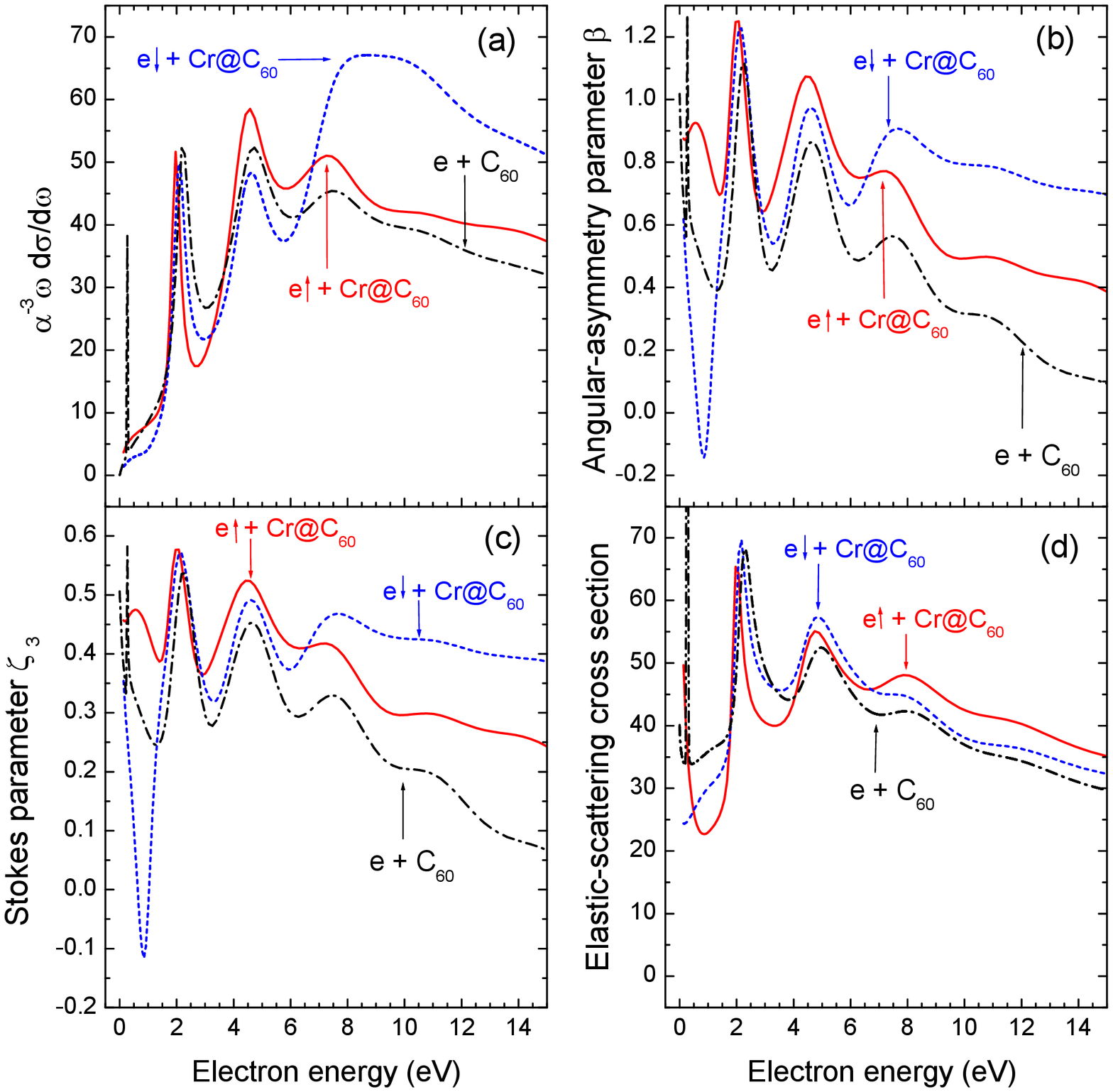}}
\caption{Calculated SPHF $\alpha^{-3} \omega\frac{d\sigma_{\uparrow(\downarrow)}}{d\omega}$ (in atomic units), angular-asymmetry parameter $\beta_{\uparrow(\downarrow)}$, and Stokes polarization parameter
$\zeta_{3}$$\uparrow$$(\downarrow)$$|_{\theta=90^{\circ}}$ of low-frequency bremsstrahlung, as well as the  total electron elastic-scattering cross section $\sigma_{\rm el\uparrow(\downarrow)}$
(in units of $20\, a_{0}^2$)
upon
electron collision of spin-up and spin-down electrons with Cr@C$_{60}$ and empty C$_{60}$, as marked.}
\label{FigCr}
\end{figure*}

\begin{figure*}[ht]
\center{\includegraphics[width=16cm]{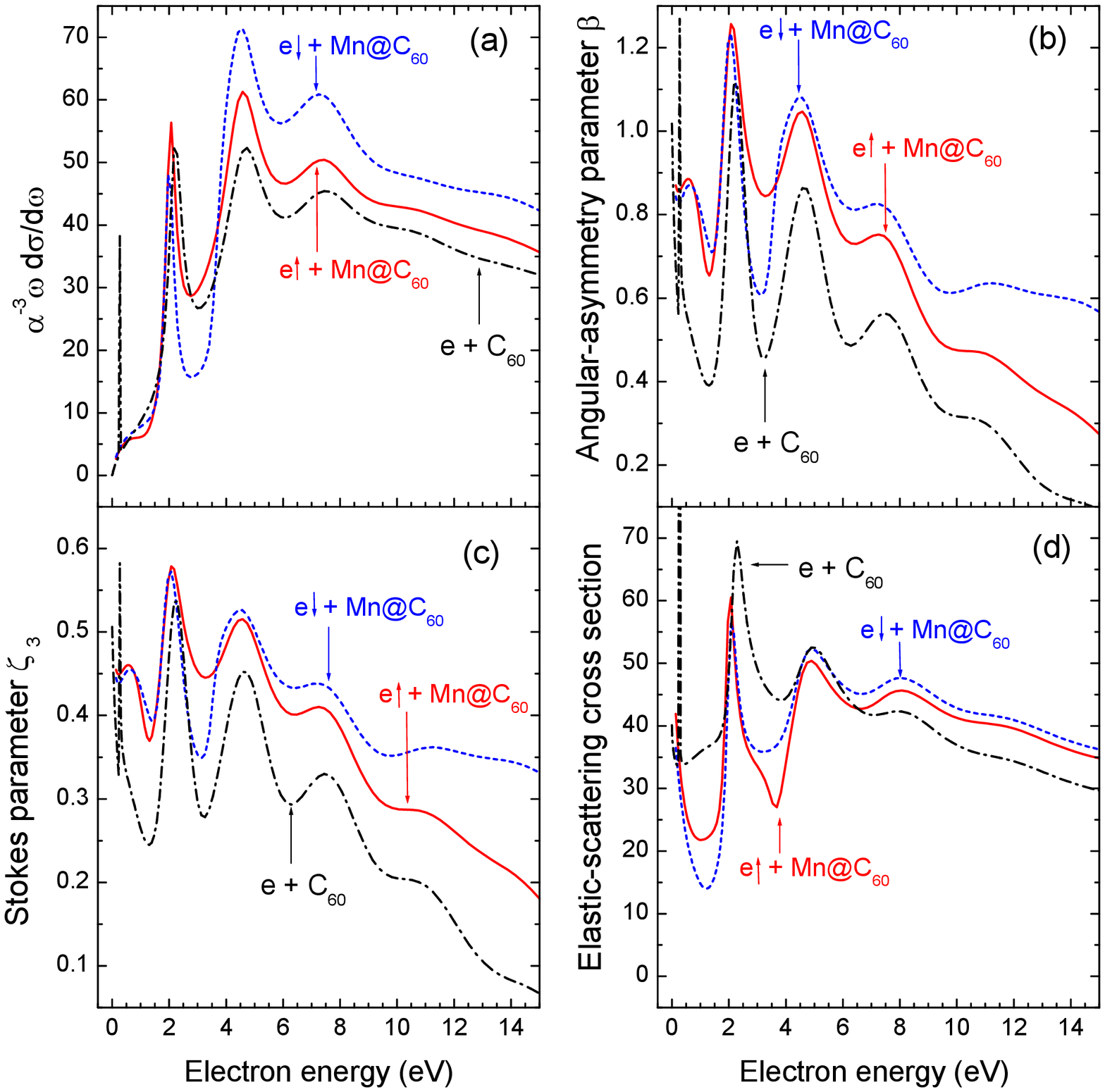}}
\caption{Calculated SPHF $\alpha^{-3} \omega\frac{d\sigma_{\uparrow(\downarrow)}}{d\omega}$ (in atomic units), angular-asymmetry parameter $\beta_{\uparrow(\downarrow)}$, and Stokes polarization parameter
$\zeta_{3}$$\uparrow$$(\downarrow)$$|_{\theta=90^{\circ}}$ of low-frequency bremsstrahlung,
as well as the  total
electron elastic-scattering cross section $\sigma_{\rm el\uparrow(\downarrow)}$ (in units of $20\, a_{0}^2$)
upon electron collision of spin-up and spin-down electrons with Mn@C$_{60}$ and empty C$_{60}$, as marked.}
\label{FigMn}
\end{figure*}

Finally, the spectral density $d\sigma_{\uparrow(\downarrow)}/d\omega$, angular-asymmetry parameter $\beta_{\uparrow(\downarrow)}$, and Stokes polarization parameter
$\zeta_{3}$$\uparrow$$(\downarrow)$$|_{\theta=90^{\circ}}$ of low-frequency bremsstrahlung, as well as the  total
electron elastic-scattering cross sections $\sigma_{\rm el\uparrow(\downarrow)}$ upon electron collision with Cr@C$_{60}$ and Mn@C$_{60}$ are depicted in Figs.~\ref{FigCr} and \ref{FigMn}, respectively.

Note how total spin-up and spin-down electron elastic-scattering cross sections
 $\sigma_{\rm el\uparrow}$ and $\sigma_{\rm el\downarrow}$ differ clearly from each other, both in the case of electron collision with Cr@C$_{60}$ and Mn@C$_{60}$. This is in contrast to
 the previous case of $e+{\rm  N@C_{60}}$ collision for the following reasons. First, because Cr and Mn have greater spins than the N atom. Second, because of a strong drain of the electron density from Cr and Mn
 to the C$_{60}$ cage, but the absence of the drain in N@C$_{60}$.

 Next, the differences between  $\sigma_{\rm el\uparrow}$ and $\sigma_{\rm el\downarrow}$  are seen to be stronger for the case of $e + {\rm Cr@C_{60}}$ than $e + {\rm Mn@C_{60}}$ collision. This, in turn,
 is (a) because the spin of Cr is greater than the spin of Mn and (b) because the C$_{60}$ cage is spin-up ``charged'' in Cr@C$_{60}$ but spin-down ``charged'' in Mn@C$_{60}$, as
   was discussed above.

 Furthermore, note that, depending on the electron energy,
 either $\sigma_{\rm el\uparrow}$ is greater than $\sigma_{\rm el\downarrow}$ or vice verse. Moreover, interestingly enough, the noted feature develops differently in $e + \rm{Cr@C_{60}}$  than
$e+ {\rm Mn@C_{60}}$ collision. Indeed, one can see that $\sigma_{\rm el\uparrow}^{\rm Cr@C_{60}} < \sigma_{\rm el\downarrow}^{\rm Cr@C_{60}}$
in the region of $\epsilon < 7$ eV (with the exception of an extremely narrow domain
$\Delta\epsilon$ near the left slope of a maximum at $\epsilon \approx 2$ eV), whereas $\sigma_{\rm el\uparrow}^{\rm Cr@C_{60}} > \sigma_{\rm el\downarrow}^{\rm Cr@C_{60}}$
 everywhere for $\epsilon  > 7$ eV. In contrast, $\sigma_{\rm el\uparrow}^{\rm Mn@C_{60}} > \sigma_{\rm el\downarrow}^{\rm Mn@C_{60}}$
in the region of $\epsilon < 2$ eV, but  $\sigma_{\rm el\uparrow}^{\rm Mn@C_{60}} < \sigma_{\rm el\downarrow}^{\rm Mn@C_{60}}$ thereafter.

Next, note that differences between the spin-up and spin-down \textit{bremsstrahlung} parameters are stronger, both quantitatively and qualitatively, than the differences between the corresponding total electron
 \textit{elastic-scattering cross sections}. For example, note how an insignificant, weakly-developed maximum in $\sigma_{\rm el\downarrow}^{\rm Cr@C_{60}}$ at $\epsilon \approx 8$ eV transforms into a strong, broad maximum with a
plateau-type top in the $d\sigma_{\downarrow}^{\rm Cr@C_{60}}/d\omega$ bremsstrahlung spectral density. The latter also differs quantitatively and qualitatively from  $d\sigma_{\downarrow}^{\rm Mn@C_{60}}/d\omega$.

Another noteworthy result is that the depicted in Fig.~\ref{FigCr} Stokes polarization parameter $\zeta_{3}^{\rm Cr@C_{60}}$$\downarrow$ changes its sign twice in the narrow region of
$\epsilon < 1.5$ eV, whereas $\zeta_{3}^{\rm Cr@C_{60}}$$\uparrow$ remains always positive. This is in contrast to $e+ {\rm Mn@C_{60}}$ collision where both
$\zeta_{3}^{\rm Mn@C_{60}}$$\uparrow$ and
$\zeta_{3}^{\rm Mn@C_{60}}$$\downarrow$ remain always positive (Fig.~\ref{FigMn}).

To summarize, one learns from the above discussion that a  drain of the electron density, especially spin-dependent drain, from an encapsulated atom to the C$_{60}$ cage has important consequences. It results in the enhanced sensitivity of $e+A@{\rm C_{60}}$ collision both to the individuality of an encapsulated atom (cp.\,N vs.\,Cr\, vs.\,Mn) and spin-polarization of an incident electron.

\subsubsection{Mo@C$_{60}$ and Tc@C$_{60}$}

The Mo(...${4\rm d}^{5}$$\uparrow$${5\rm s}^{1}$$\uparrow$,\,$^{7}$S) and Tc(...${4\rm d}^{5}$$\uparrow$${5\rm s}^{1}$$\uparrow$${5\rm s}^{1}$$\uparrow$,\,$^{6}$S) atoms have the valence-shell structures
which look similar to those of  the Cr(...${3\rm d}^{5}$$\uparrow$${4\rm s}^{1}$$\uparrow$,\,$^{7}$S)  and Mn(...${3\rm d}^{5}$$\uparrow$${4\rm s}^{1}$$\uparrow$$4s^{1}$$\downarrow$,\,$^{6}$S), respectively. However,
 because of the greater values of the principal quantum numbers, the ${4\rm d}$$\uparrow$- and ${5\rm s}$$\uparrow$($\downarrow$)-wavefunctions of Mo and Tc have an additional node compared to the
 corresponding wavefunctions of Cr and Mn.  It is,
therefore, interesting to learn about differences and/or similarities between electron elastic scattering and bremsstrahlung on  Mo@C$_{60}$ versus Cr@C$_{60}$, as well as Mn@C$_{60}$ versus Tc@C$_{60}$.

\begin{figure}[ht]
\center{\includegraphics[width=8cm]{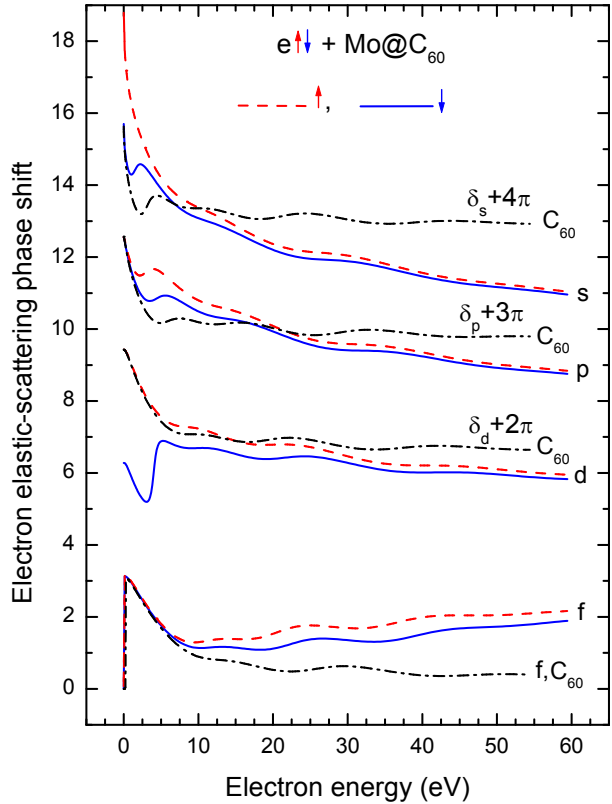}}
\caption{Calculated SPHF electron elastic-scattering phase shifts (in units of radian) $\delta_{\ell\uparrow}^{\rm Mo@C_{60}}$ and $\delta_{\ell\downarrow}^{\rm Mo@C_{60}}$  ($\ell \le 3$) upon scattering
of spin-up and spin-down electrons off Mo@C$_{60}$, as well as $\delta_{\ell}^{\rm C_{60}}$ due to electron collision with empty C$_{60}$, as marked.}
\label{FigMo03}
\end{figure}

\begin{figure}[ht]
\center{\includegraphics[width=8cm]{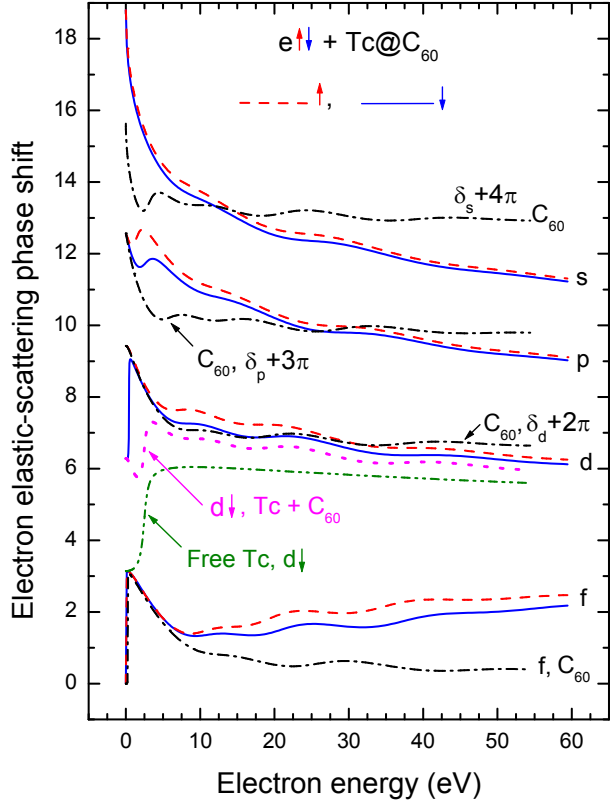}}
\caption{ Calculated SPHF electron elastic scattering phase shifts (in units of radian) $\delta_{\ell\uparrow}^{\rm Tc@C_{60}}$ and $\delta_{\ell\downarrow}^{\rm Tc@C_{60}}$  ($\ell \le 3$) upon scattering
of spin-up and spin-down electrons off Tc@C$_{60}$, as well as $\delta_{\ell}^{\rm C_{60}}$ due to electron collision with empty C$_{60}$, as marked. Also plotted are the free-Tc $\delta_{d\downarrow}^{\rm Tc}$
phase shift (dash-dot-dotted line) and ${\tilde \delta}_{d\downarrow}^{\rm Tc} = \delta_{d\downarrow}^{\rm Tc} + \delta_{d}^{\rm C_{60}}$ (dotted line).}
\label{FigTc03}
\end{figure}

Calculated SPHF $s$-, $p$-, $d$-, and $f$-electron elastic-scattering phase shifts of spin-up and spin-down electrons collided with Mo@C$_{60}$ and Tc@C$_{60}$ are depicted in Figs.~\ref{FigMo03} and \ref{FigTc03},
respectively.

Looking at the values of these phase shifts at $\epsilon = 0$,  one learns about binding properties of the Mo@C$_{60}$ and Tc@C$_{60}$ fullerenes. They appear to be the same as those of Cr@C$_{60}$ and Mn@C$_{60}$.
Namely, both Mo@C$_{60}$ and Tc@C$_{60}$ are capable of binding an extra electron into a $s$-, or $p$-, or $d$-bound state, but not into a state with $\ell \ge 3$, regardless of the spin polarization
of the electron. This, again, is in contrast to the Ba@C$_{60}$ case.

Next, note that the $\delta_{\ell\uparrow}^{\rm Mo@C_{60}}$ and $\delta_{\ell\uparrow}^{\rm Tc@C_{60}}$ phase shifts, except for $\delta_{d\downarrow}^{\rm Tc@C_{60}}$, behave similarly to the corresponding phase shifts  due to electron scattering off Cr@C$_{60}$ and Mn@C$_{60}$ (Figs.~\ref{FigCr03} and \ref{FigMn03}). The following two cases of the similarities and differences between these phase shifts are particularly noteworthy.

First, one can see that the spin-down phase shift $\delta_{s\downarrow}^{\rm Mo@C_{60}}$ (similar to $\delta_{s\downarrow}^{\rm Cr@C_{60}}$) has a well-developed oscillation near threshold. This is in contrast to the spin-up phase shift $\delta_{s\uparrow}^{\rm Mo@C_{60}}$ (similar to $\delta_{s\uparrow}^{\rm Cr@C_{60}}$) as well as the $s$-phase shifts $\delta_{s\uparrow(\downarrow)}^{\rm Tc@C_{60}}$
(similar to $\delta_{s\uparrow(\downarrow)}^{\rm Mn@C_{60}}$).
The noted feature admits the same interpretation developed for the $s$-phase shifts upon electron collision with Cr@C$_{60}$ and Mn@C$_{60}$ (refer to the discussion above).
Accordingly, the oscillation in
 $\delta_{s\downarrow}^{\rm Mo@C_{60}}$ is because of the constructive three-wave interference between the incident $s$$\downarrow$-electronic wave and electronic waves scattered off the Mo atom and C$_{60}$, in contrast to the case of
 $s$$\uparrow$-scattering.

Second, an interesting feature can be revealed if one inter-compares calculated $d$$\downarrow$-phase shifts $\delta_{d\downarrow}^{\rm Tc@C_{60}}$ (Fig.~\ref{FigTc03}),  $\delta_{d\downarrow}^{\rm Mo@C_{60}}$
(Fig.~\ref{FigMo03}), $\delta_{d\downarrow}^{\rm Mn@C_{60}}$ (Fig.~\ref{FigMn03}), and $\delta_{d\downarrow}^{\rm Cr@C_{60}}$ (Fig.~\ref{FigCr03}) between themselves. Namely, one finds that
$\delta_{d\downarrow}^{\rm Tc@C_{60}}$ has a sharp \textit{maximum} near threshold in contrast to the \textit{minima} in all other phase shifts under discussion. The origin of the minima in
$\delta_{d\downarrow}^{\rm Mn@C_{60}}$ and $\delta_{d\downarrow}^{\rm Cr@C_{60}}$ was interpreted above. A trial calculation
showed that the origin of the minimum in $\delta_{d\downarrow}^{\rm Mo@C_{60}}$ admits exactly the same interpretation as well. Namely, in accordance with Levinson theorem the free-Mo phase shift drops down to the value of
$\pi$, i.e., $\delta_{d\downarrow}^{\rm Mo}(0) \rightarrow \pi$, at $\epsilon \rightarrow 0$. Then, with the help of the independent-scattering approximation, one easily finds that $\delta_{d\downarrow}^{\rm Mo@C_{60}}$, defined  as $\delta_{d\downarrow}^{\rm Mo@C_{60}} \approx \delta_{d\downarrow}^{\rm Mo} + \delta_{d}^{\rm C_{60}}$, must have a minimum near threshold. An attempt to explain the emergence of the \textit{maximum}
 in $\delta_{d\downarrow}^{\rm Tc@C_{60}}$ near threshold in the similar manner, however, basically fails. Indeed, the addition of the free-Tc phase shift (Fig.~\ref{FigTc03}, dash-dot-dotted line) to  $\delta_{d}^{\rm C_{60}}$ results in the emergence of a minimum rather than a maximum in $\delta_{d\downarrow}^{\rm Tc} + \delta_{d}^{\rm C_{60}}$ (Fig.~\ref{FigTc03}, dotted line). The implication is that the
 independent-scattering approximation is a poor approximation for the case of the $d$$\downarrow$-electron scattering off Tc@C$_{60}$, at low electron energies. Nevertheless, the maximum in
 $\delta_{d\downarrow}^{\rm Tc@C_{60}}$ seems to have everything to do with the behavior of the free-Tc phase shift $\delta_{d\downarrow}^{\rm Tc}$ at $\epsilon \rightarrow 0$. Otherwise, it would be totally
 unclear why there is a sudden decrease of  $\delta_{d\downarrow}^{\rm Tc@C_{60}}$ to $\delta_{d\downarrow}^{\rm Tc@C_{60}}(0)=2\pi$ (note, $2\pi=[\delta_{d\downarrow}^{\rm Tc}(0)=\pi] + [\delta_{d}^{\rm C_{60}}(0)=\pi]$).

\begin{figure}[ht]
\center{\includegraphics[width=8cm]{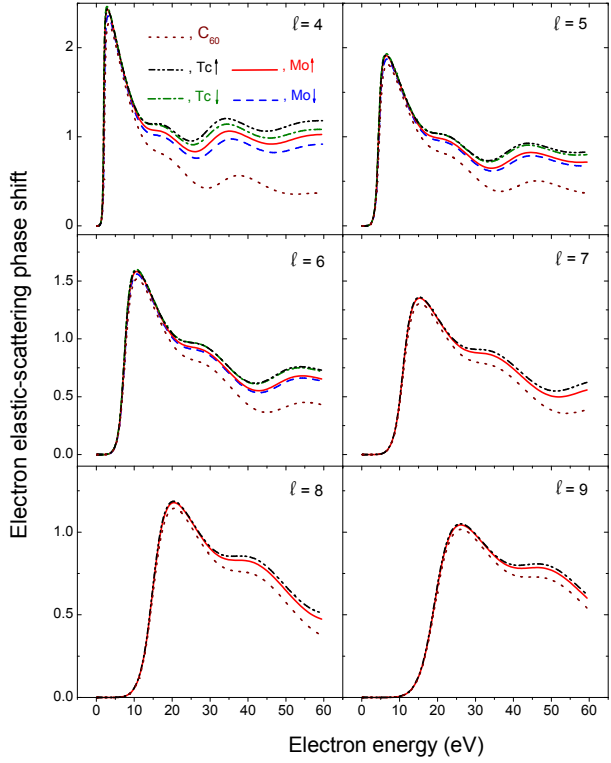}}
\caption{Calculated SPHF  electron elastic-scattering phase shifts (in units of radian)  upon scattering
of spin-up and spin-down electrons ($4 \le \ell \le 9$) off Mo@C$_{60}$, Tc@C$_{60}$, and C$_{60}$, as marked. The corresponding phase shifts due to electron scattering from empty C$_{60}$ are depicted as well.}
\label{FigMoTc49}
\end{figure}

Next, calculated SPHF electron elastic-scattering phase shifts of spin-up and spin-down electrons with $\ell \ge 4$ upon their collision with Mo@C$_{60}$ and Tc@C$_{60}$ are depicted in  Fig.~\ref{FigMoTc49}.

One can see that, similar to the case of electron scattering off Cr@C$_{60}$ and Mn@C$_{60}$, the difference between the spin-up and spin-down phase shifts is quickly decreasing with increasing $\ell$ and practically
vanishes. The differences
in the phase shifts between electron scattering off Mo@C$_{60}$ and C$_{60}$, as well as between Tc@C$_{60}$ and C$_{60}$ are decreasing as well, with increasing $\ell$. Yet, they remain visible for all $\ell$s
under discussion.

\begin{figure*}[ht]
\center{\includegraphics[width=16cm]{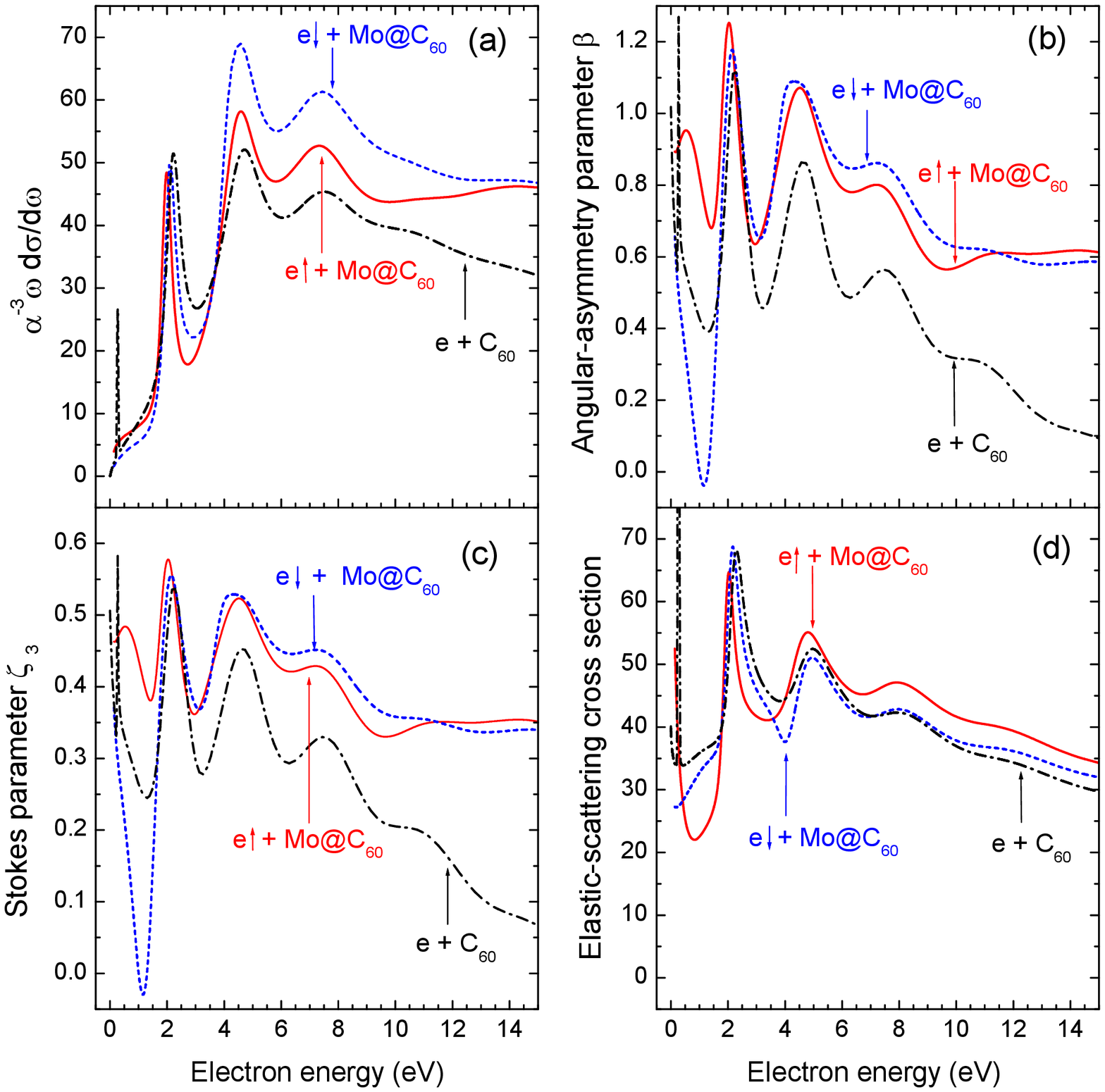}}
\caption{Calculated SPHF $\alpha^{-3} \omega\frac{d\sigma_{\uparrow(\downarrow)}}{d\omega}$ (in atomic units), angular-asymmetry parameter $\beta_{\uparrow(\downarrow)}$, and Stokes polarization parameter
$\zeta_{3}$$\uparrow$$(\downarrow)$$|_{\theta=90^{\circ}}$ of low-frequency bremsstrahlung, as well as the  total electron elastic-scattering cross section  $\sigma_{\rm el\uparrow(\downarrow)}$
(in units of $20\,a_{0}^2$) upon
electron collision of spin-up and spin-down electrons with Mo@C$_{60}$ and empty C$_{60}$, as marked.}
\label{FigMo}
\end{figure*}

\begin{figure*}[ht]
\center{\includegraphics[width=16cm]{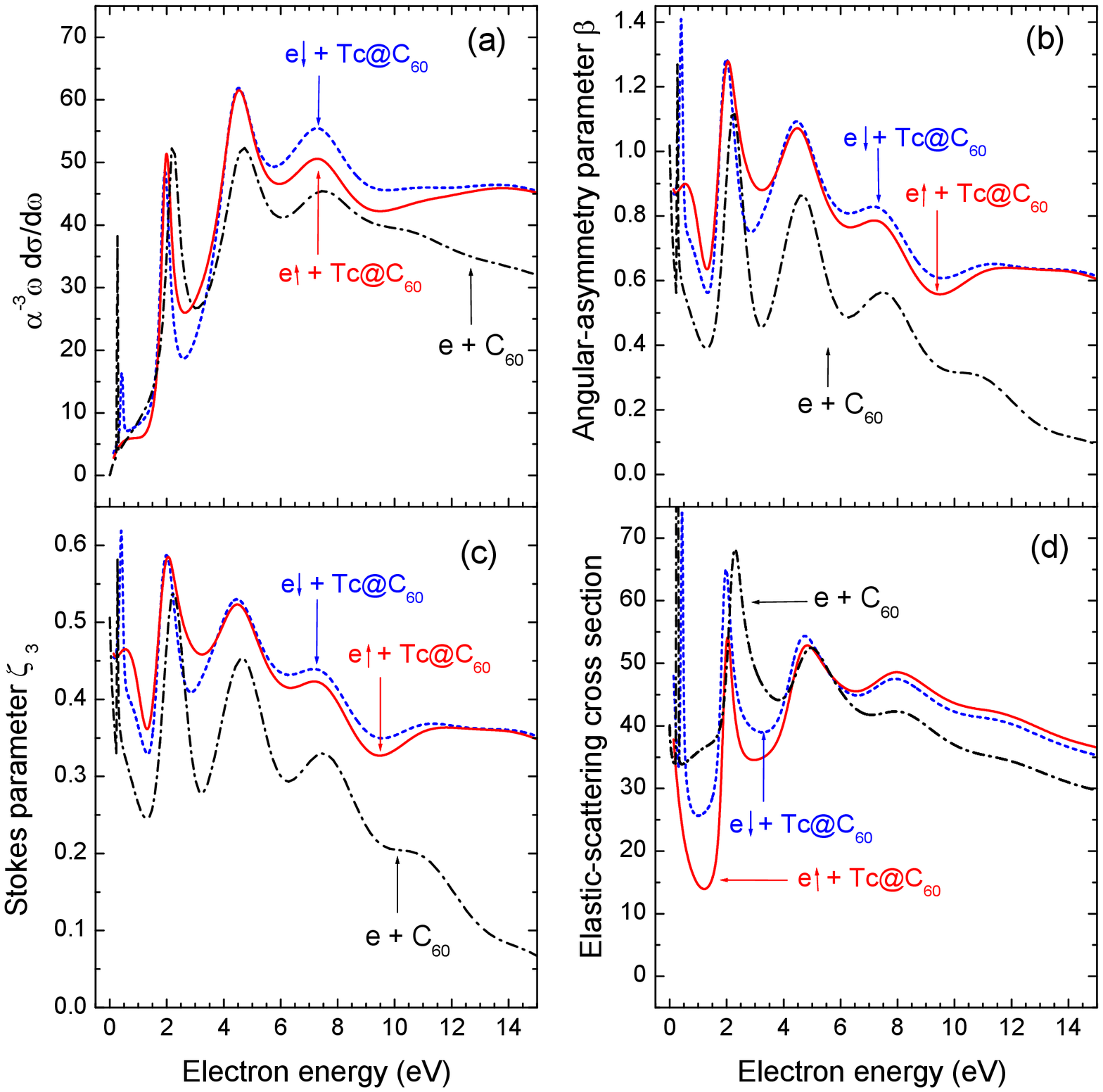}}
\caption{Calculated SPHF $\alpha^{-3} \omega\frac{d\sigma_{\uparrow(\downarrow)}}{d\omega}$ (in atomic units), angular-asymmetry parameter $\beta_{\uparrow(\downarrow)}$, and Stokes polarization parameter
$\zeta_{3}$$\uparrow$$(\downarrow)$$|_{\theta=90^{\circ}}$ of low-frequency bremsstrahlung,
as well as the  total
electron elastic-scattering cross section $\sigma_{\rm el\uparrow(\downarrow)}$ (in units of $20\,a_{0}^2$) due to electron collision of spin-up and spin-down electrons with Tc@C$_{60}$ and empty C$_{60}$, as marked.}
\label{FigTc}
\end{figure*}

Finally, the spectral density $d\sigma_{\uparrow(\downarrow)}/d\omega$, angular-asymmetry parameter $\beta_{\uparrow(\downarrow)}$, and Stokes polarization-parameter
$\zeta_{3}$$\uparrow$$(\downarrow)$$|_{\theta=90^{\circ}}$ of low-frequency bremsstrahlung, as well as the  total
electron elastic-scattering cross sections $\sigma_{\rm el\uparrow(\downarrow)}$ upon electron collision with Mo@C$_{60}$ and Tc@C$_{60}$ are depicted in Figs.~\ref{FigMo} and \ref{FigTc}, respectively.

One can see that the differences between spin-up and spin-down electron elastic scattering as well as bremsstrahlung are greater for $e + {\rm Mo@C_{60}}$ than $e + {\rm Tc@C_{60}}$ collision.
This is exactly for the same reason as the differences between electron collisions with Cr@C$_{60}$ and Mn@C$_{60}$. This is not accidental, because Mo (similar to Cr vs.\,Mn), has a greater spin than
Tc. In addition, Mo (similar to Cr vs.\,Mn) ``charges'' the C$_{60}$ shell by the spin-up electron density whereas Tc does the exact opposite, as was discussed earlier in the paper.

\begin{figure*}[ht]
\center{\includegraphics[width=16cm]{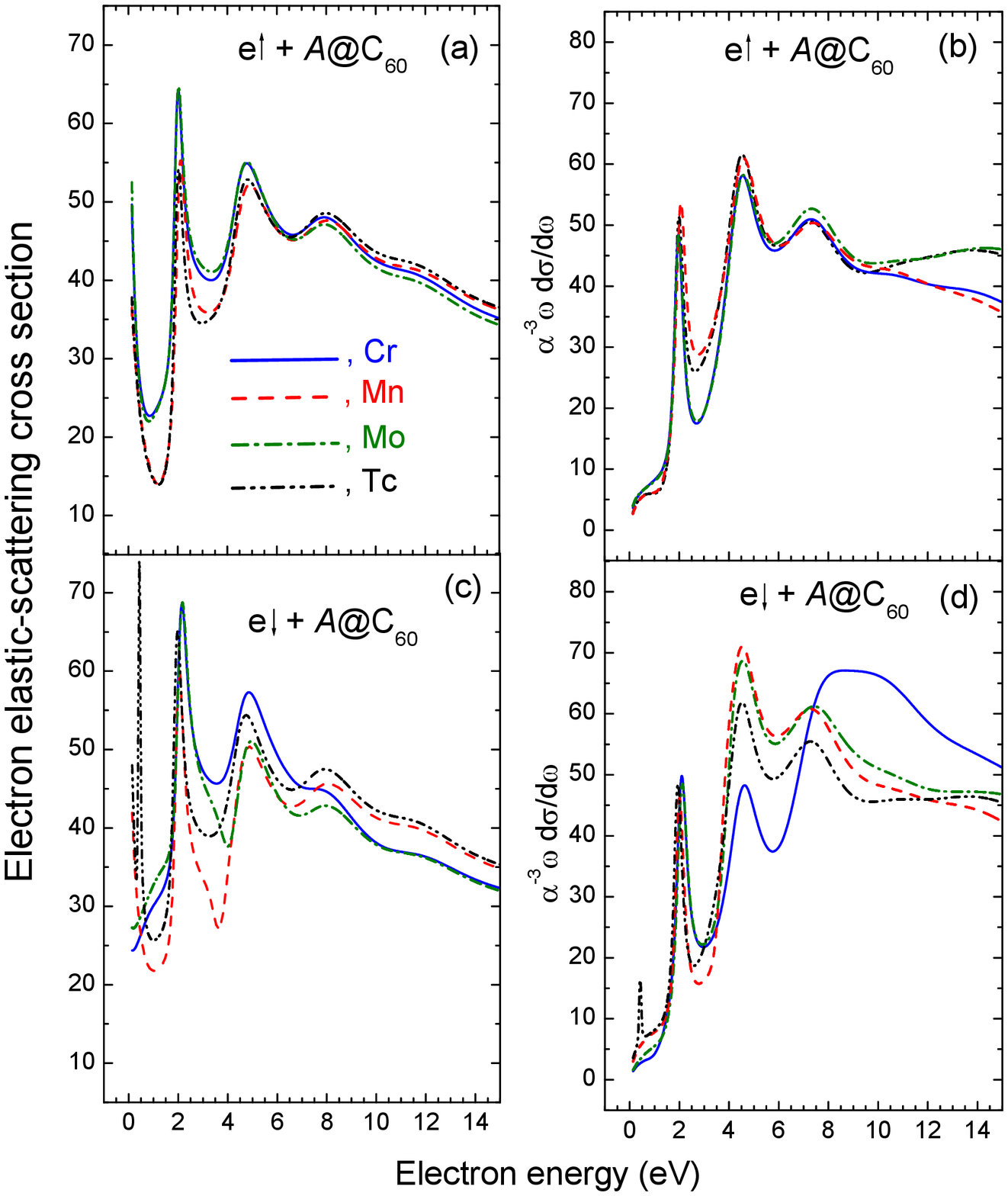}}
\caption{Calculated SPHF electron low-frequency bremsstrahlung spectral density $\alpha^{-3}\omega \frac{d\sigma_{\uparrow(\downarrow)}}{d\omega}$ (in atomic units) and
 the total electron elastic-scattering cross section $\sigma_{\rm el\uparrow(\downarrow)}$ (in units of $20\,a_{0}^{2}$)
 upon collision with Cr@C$_{60}$ (solid line), Mn@C$_{60}$ (dashed line), Mo@C$_{60}$ (dash-dotted line), and Tc@C$_{60}$ (dash-dot-dotted line), as marked.}
\label{CrMnMoTc}
\end{figure*}

In view of the above, it is also interesting to compare directly calculated data for $e + {\rm Cr@C_{60}}$ and $e + {\rm Mn@C_{60}}$ scattering with those for $e + {\rm Mo@C_{60}}$ and $e + {\rm Tc@C_{60}}$
scattering, respectively (see Fig.~\ref{CrMnMoTc}).

By studying Fig.~\ref{CrMnMoTc}, one arrives at spectacular findings which are interesting in themselves.

First (see Fig.~\ref{CrMnMoTc},\,$a$), it appears that \textit{spin-up} electron \textit{elastic-scattering cross sections} for the corresponding pairs
of fullerenes with like electron configurations (Cr vs.\,Mo and Mn vs.\,Tc) are equal in the whole range of considered electron energies, to almost an excellent approximation, i.e.,
$\sigma_{\rm el\uparrow}^{\rm Cr@C_{60}} = \sigma_{\rm el\uparrow}^{\rm Mo@C_{60}}$ and $\sigma_{\rm el\uparrow}^{\rm Mn@C_{60}} = \sigma_{\rm el\uparrow}^{\rm Tc@C_{60}}$.
The same is observed for the corresponding \textit{bremsstrahlung spectral densities} as well (see Fig.~\ref{CrMnMoTc},\,$b$), though the latter is true only in the electron energy range up to approximately $6$ eV. There, indeed,
$d\sigma_{\uparrow}^{\rm Cr@C_{60}}/d\omega = d\sigma_{\uparrow}^{\rm Mo@C_{60}}/d\omega$
and $d\sigma_{\uparrow}^{\rm Mn@C_{60}}/d\omega = d\sigma_{\uparrow}^{\rm Tc@C_{60}}/d\omega$, to a nearly excellent approximation. It is only
above $\epsilon \approx 6$ eV that the corresponding bremsstrahlung spectral densities starts  deviating from each other.

Second, above approximately $8$ eV, a sort of switching occurs between the pairs of fullerenes for which  \textit{spin-up} bremsstrahlung spectral densities become nearly perfectly equaled. They now
``pair'' by a principle of the \textit{closest} proximity in the same \textit{row} of the periodic table:
$d\sigma_{\uparrow}^{\rm Cr@C_{60}}/d\omega = d\sigma_{\uparrow}^{\rm Mn@C_{60}}/d\omega$ and
$d\sigma_{\uparrow}^{\rm Mo@C_{60}}/d\omega = d\sigma_{\uparrow}^{\rm Tc@C_{60}}/d\omega$.

As for collision of \textit{spin-down} electrons with the discussed fullerenes (see Figs.~\ref{CrMnMoTc},\,$c$ and $d$), the
situation there is not so deterministic or simple as for scattering of \textit{spin-up} electrons. In particular, the case of the spin-down electron bremsstrahlung off the Cr@C$_{60}$ fullerene singles out as a stand alone case.

\subsubsection{Eu@C$_{60}$}

Eu(...${4\rm f}^{7}$$\uparrow$${6\rm s}^{1}$$\uparrow$$6s^{1}$$\downarrow$, $^{8}$S) is the first atom in the periodic table with the most capacious semifilled subshell - the $4f^{7}$$\uparrow$ subshell. The Eu atom, thus, is an atom with the highest spin. Therefore, electron collision with Eu@C$_{60}$ represents another interesting case study.

\begin{figure}[ht]
\center{\includegraphics[width=8cm]{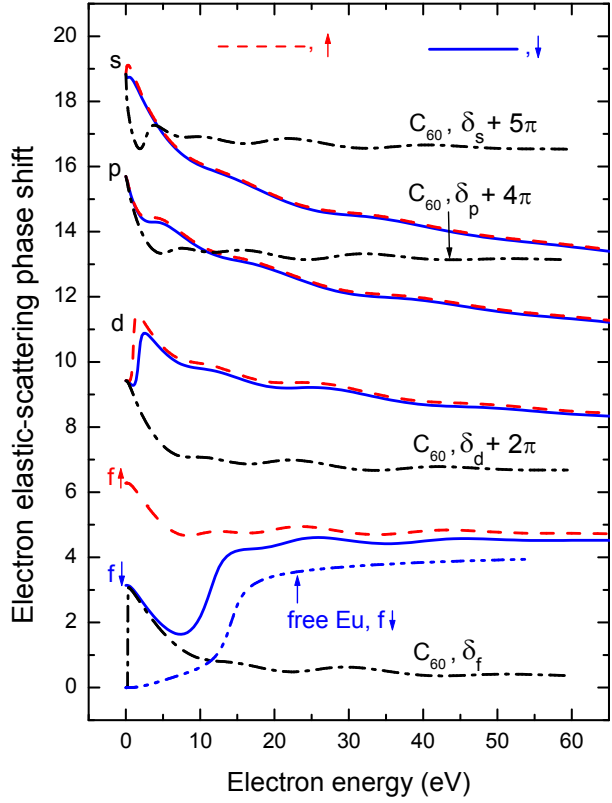}}
\caption{Calculated SPHF electron elastic-scattering phase shifts (in units of radian) $\delta_{\ell\uparrow(\downarrow)}^{\rm Eu@C_{60}}$
of incident spin-down (solid line) and spin-up (dashed line) electrons with $\ell \leq 3$ scattered off Eu@C$_{60}$, the free-atom $f$-phase shifts $\delta_{f\downarrow}^{\rm Eu}$  of spin-down electrons scattered off free Eu (dash-dot-dotted line), and the phase shifts $\delta_{\ell}^{\rm C_{60}}$ (dash-dotted line) upon electron scattering off empty C$_{60}$.}
\label{FigEu03}
\end{figure}

Calculated SPHF electron elastic-scattering phase shifts of spin-up and spin-down electrons with $\ell \leq 3$ scattered off Eu@C$_{60}$ are depicted in Fig.~\ref{FigEu03}.

First, looking at the values of the depicted phase shifts at $\epsilon=0$, one concludes, with the help of Levinson theorem, that, similar to Ba@C$_{60}$, the Eu@C$_{60}$ fullerene  is capable of binding an extra electron into a $p$-, or $d$-, or $f$-state, but not into a $s$-state.

Second, note how the $f$-phase shift $\delta_{f\uparrow}^{\rm Eu@C_{60}}$ of a scattered \textit{spin-up} electron differs drastically from the $f$-phase shift $\delta_{f\downarrow}^{\rm Eu@C_{60}}$ of a scattered \textit{spin-down} electron. The reason for this is similar to the reasons for the drastic differences between the $d$$\uparrow$- and $d$$\downarrow$-phase shifts in electron collisions with the transition-atom metallo-fullerenes discussed above. Specifically, this is owing to the presence of the only spin-unpaired $f$-subshell ($4f^{7}$$\uparrow$) in the ground state of Eu.
In accordance with Levinson theorem, the free-Eu spin-down phase shift $\delta_{f\downarrow}^{\rm Eu} \rightarrow 0$ at $\epsilon \rightarrow 0$. Therefore, the $\delta_{f\downarrow}^{\rm Eu}$ phase shift
(Fig.~\ref{FigEu03}, dash-dot-dotted dotted line) is decreasing with decreasing energy $\epsilon$. Starting at $\epsilon \approx 15$ eV, the rate of the decrease of $\delta_{f\downarrow}^{\rm Eu@C_{60}}$ overtakes the rate of the increase of
$\delta_{f}^{\rm C_{60}}$, with decreasing $\epsilon$. Correspondingly, in accordance with the independent-scattering approximation,  Eq.~(\ref{EqD+D}), the phase shift
$\delta_{f\downarrow}^{\rm Eu@C_{60}} \approx \delta_{f\downarrow}^{\rm Eu} + \delta_{f\downarrow}^{\rm C_{60}}$ starts falling down, with decreasing $\epsilon$.
However, at yet lower energies, below approximately $8$ eV, the situation reverts to the exact opposite. It is now the rate of the increase of
$\delta_{f}^{\rm C_{60}}$ that overtakes
 the rate of the decrease of $\delta_{f\downarrow}^{\rm Eu}$. As a result, in accordance with the independent-scattering approximate, the phase shift
$\delta_{f\downarrow}^{\rm Eu@C_{60}}$ starts increasing at $\epsilon \rightarrow 0$. Correspondingly, the well-developed low-energy minimum emerges in $\delta_{f\downarrow}^{\rm Eu@C_{60}}$,
depicted in Fig.~\ref{FigEu03}. In contrast to $\delta_{f\downarrow}^{\rm Eu}$, the free-Eu phase shift $\delta_{f\uparrow}^{\rm Eu}$ (not plotted) steadily increases to the value of $\pi$ at $\epsilon \rightarrow 0$
(the latter is due to Levinson theorem). Accordingly, so does the phase shift $\delta_{f\uparrow}^{\rm Eu@C_{60}}$ as well. As a result, the $\delta_{f\downarrow}^{\rm Eu@C_{60}}(\epsilon)$ and $\delta_{f\uparrow}^{\rm Eu@C_{60}}$ phase shifts take drastically different routes at low electron energies (see Fig.~\ref{FigEu03}).

Third, the differences between other phase shifts with $\ell \neq f$, $\delta_{(\ell \neq f)}^{\rm Eu@C_{60}}$, are seen to be small to negligible. This is because the $4f^{7}$$\uparrow$ subshell of Eu is collapsed
deep into the inner region of the atom. Consequently, exchange interaction between incident spin-up (spin-down) electrons, whose $\ell \neq f$, and the $4f$$\uparrow$-electrons is small. This largely eliminates differences in the overall impact of exchange interaction
on $\delta_{(\ell\neq f)\uparrow}^{\rm Eu@C_{60}}$  and $\delta_{(\ell\neq f)\downarrow}^{\rm Eu@C_{60}}$, thereby making them nearly equal.

\begin{figure}[ht]
\center{\includegraphics[width=8cm]{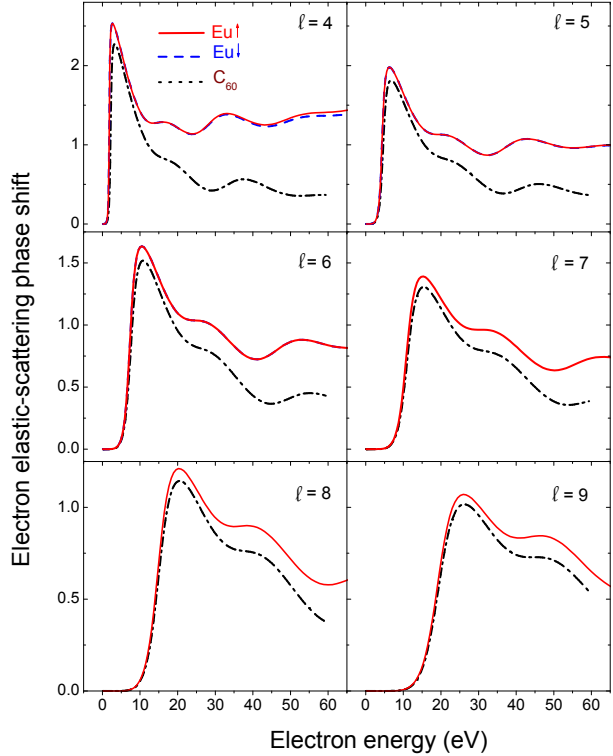}}
\caption{Calculated SPHF electron elastic-scattering phase shifts (in units of radian) $\delta_{\ell\uparrow(\downarrow)}^{\rm Eu@C_{60}}$  and $\delta_{\ell}^{\rm C_{60}}$
upon collision of incident spin-up (spin-down) electrons, whose $\ell \geq 4$, with Eu@C$_{60}$ and empty C$_{60}$, as marked.}
\label{FigEu49}
\end{figure}

Calculated SPHF phase shifts $\delta_{(\ell\ge 4)\uparrow(\downarrow)}^{\rm Eu@C_{60}}$ of spin-up and spin-down electronic waves for $\ell \geq 4$ are plotted in Fig.~\ref{FigEu49}.

One can see that there is little-to-no spin-dependence of the phase shifts with $\ell \ge 4$. This is in accordance with the explanation provided above.

Next, note that, for a given $\ell=\ell_{0}$, the difference between $\delta_{\ell_{0}\uparrow(\downarrow)}^{\rm Eu@C_{60}}$ and the phase shift $\delta_{\ell_{0}}^{\rm C_{60}}$  is increasing with increasing $\epsilon$. This is in line with the corresponding results found above for electron scattering off other fullerenes.

Furthermore, note how the difference
between $\delta_{(\ell \ge 4)\uparrow(\downarrow)}^{\rm Eu@C_{60}}$ and $\delta_{\ell \ge 4}^{\rm C_{60}}$ is decreasing with increasing $\ell$. Yet, the difference  remains quite visible for all $\ell$s, similar, e.g.,
 to the cases of $e + {\rm Ba@C_{60}}$ and $e +{\rm Tc@C_{60}}$ scattering.

\begin{figure*}[ht]
\center{\includegraphics[width=16cm]{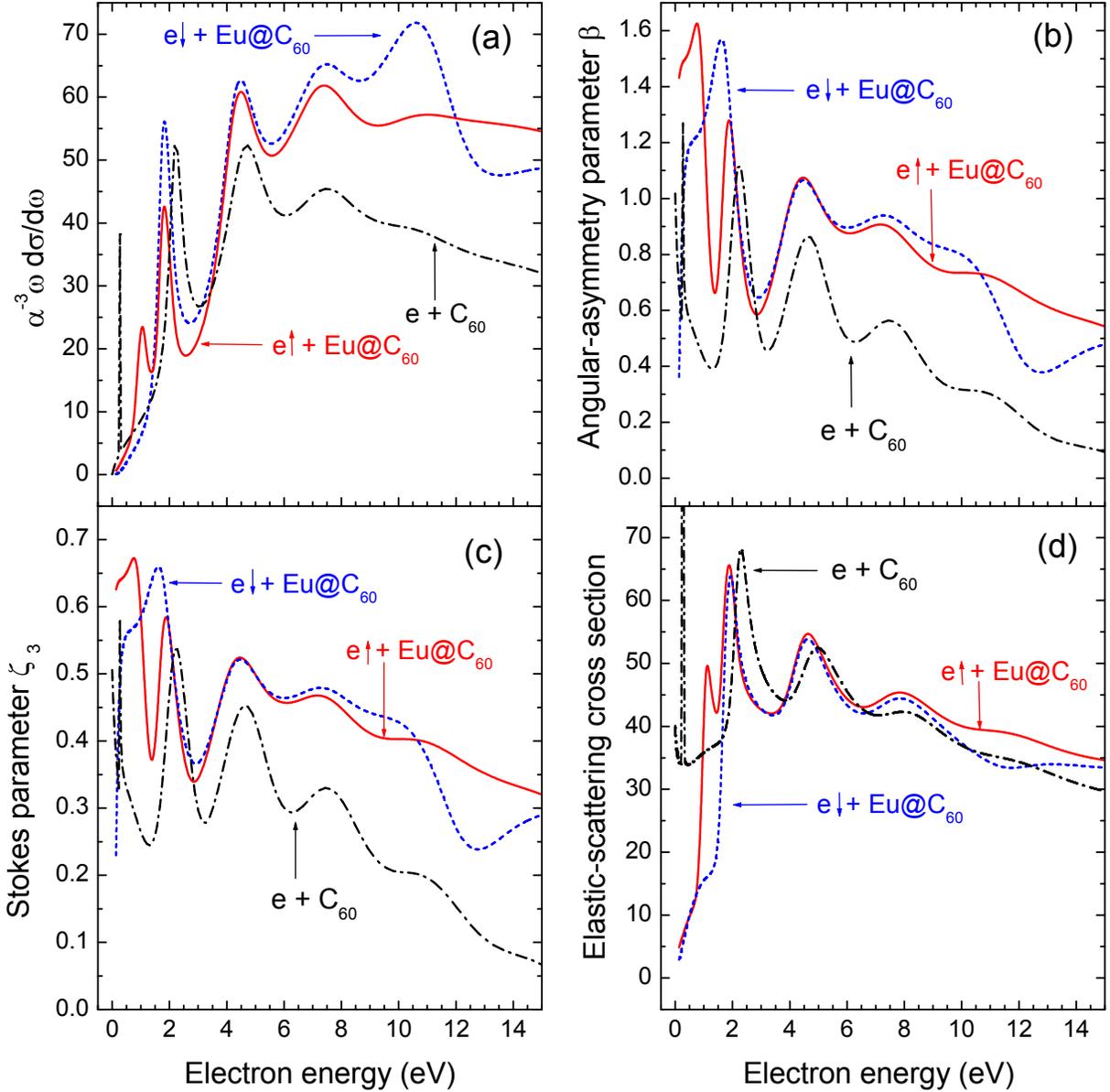}}
\caption{Calculated data for (a) electron low-frequency bremsstrahlung $\alpha^{-3}\omega \frac{d\sigma_{\uparrow(\downarrow)}}{d\omega}$, (b) angular asymmetry parameter $\beta_{\uparrow(\downarrow)}$, (c) Stokes polarization parameter $\zeta_{3}$$\uparrow$$(\downarrow)$$|_{\theta=90^{\circ}}$, and
 (d) total electron elastic-scattering cross section $\sigma_{\rm el\uparrow(\downarrow)}$ (in units of $20\,a_{0}$) for spin-up and spin-down electron collisions with Eu@C$_{60}$ and empty C$_{60}$, as marked.}
\label{FigEu}
\end{figure*}

Finally, calculated SPHF spin-up and spin-down spectral density $d\sigma_{\uparrow(\downarrow)}/d\omega$, angular-asymmetry parameter $\beta_{\uparrow(\downarrow)}$, and Stokes polarization parameter $\zeta_{3}$$\uparrow$$(\downarrow)$$|_{\theta=90^{\circ}}$ of low-frequency bremsstrahlung, as well as the  total
electron elastic-scattering cross sections $\sigma_{\rm el\uparrow(\downarrow)}$ due to electron scattering off Eu@C$_{60}$ are depicted in Fig.~\ref{FigEu}.

One can see that there are energy regions where the electron elastic-scattering cross sections, as well the parameters of bremsstrahlung for spin-up electrons differ either little-to-none or strongly from the corresponding results for spin-down electrons. This, once again, is similar to results of the other case studies discussed above.

However, what makes the Eu@C$_{60}$ additionally exclusive, compared to the discussed other fullerenes, is that both \textit{spin-up}  $d\sigma_{\uparrow}^{\rm Eu@C_{60}}/d\omega$ and
$\sigma_{\rm el\uparrow}^{\rm Eu@C_{60}}$ have a well-developed narrow maximum at $\epsilon \approx 1$ eV. The latter is absent in the spectra of the other fullerenes. A trial calculation showed that this maximum is due to the incident $d$$\uparrow$-electronic wave. This is indicative of the entrapment of the $d$$\uparrow$-electronic wave inside Eu@C$_{60}$, at $\epsilon \approx 1$ eV. Interesting, the maximum is absent in
\textit{spin-down} $d\sigma_{\downarrow}^{\rm Eu@C_{60}}/d\omega$ of Eu@C$_{60}$.  This observations is interesting. It shows that the presence
of a bigger-sized soft atom $A$ inside C$_{60}$ can result in a selective (with respect to the electronic spin and/or orbital momentum $\ell$) entrapment of incident electronic waves inside $A$@C$_{60}$, at certain energies.

\section{Conclusion}

The present work has provided the detailed insight into possible features of low-energy electron elastic scattering and low-frequency bremsstrahlung upon electron collisions with $A$@C$_{60}$ fullerenes
gained in the framework of the simple and yet reasonable model static approximation. This was achieved by studying the dependence
of these processes on the individuality of encapsulated atoms $A$ and spin-polarization of incident electrons. The chosen atoms $A$ were thoughtfully picked out as the typical representatives of atoms of different rows of the periodic system. The study has also revealed modifications in binding properties of $A$@C$_{60}$ fullerenes, as well as their ability to trap (selectively) incident electronic waves, versus the size of the encaged atom. Of certain
interest is the proposed independent-scattering approximation according to which electron elastic-scattering phase shifts upon $e + A@{\rm C_{60}}$ collision can be evaluated as a simple sum of the phase shift
due to $e + \rm C_{60}$ collision and the phase shift due to collision with the free atom $A$, for certain occasions.  Results of the work identify,
at the given level of approximation, the most interesting and/or useful future measurements or more rigorous calculations to perform. The present study
also provides researchers with a wealth of the background information which is useful for future studies aimed at elucidating of the significance of dynamical polarization and correlation effects in these processes.  The authors hope that results of the present work will serve as the impetus for such studies.

\section{Acknowledgement}

This work was supported by NSF Grant no. PHY-1305085.

\section*{References}

\end{document}